\documentclass[10pt]{article}
\pdfoutput=1
\usepackage{amssymb}
\usepackage[numbers,sort&compress]{natbib}
\usepackage{amsmath,amsfonts,graphicx,epsfig}
\usepackage{bm}

\def\bea{\begin{eqnarray}}
\def\eea{\end{eqnarray}}
\def\be{\begin{equation}}
\def\ee{\end{equation}}
\def\ba{\begin{array}}
\def\ea{\end{array}}

\begin{document}

\setlength\arraycolsep{2pt}

\renewcommand{\theequation}{\arabic{section}.\arabic{equation}}
\setcounter{page}{1}

\setlength\arraycolsep{2pt}

\begin{titlepage}

\begin{center}

\vskip 1.0 cm

{\LARGE  \bf On the importance of heavy fields during inflation}

\vskip 1.0cm

{\large
Sebasti\'an C\'espedes$^{a}$, Vicente Atal$^{b}$ and Gonzalo A. Palma$^{a}$ 
}

\vskip 0.5cm

{\it

$^{a}$Physics Department, FCFM, Universidad de Chile \mbox{Av. Blanco Encalada 2008, Santiago, Chile}

$^{b}$Instituut-Lorentz for Theoretical Physics, Universiteit Leiden \mbox{Niels Bohrweg 2, Leiden, The Netherlands}

}

\vskip 1.5cm

\end{center}

\begin{abstract}

We study the dynamics of two-field models of inflation characterized by a hierarchy of masses between curvature and isocurvature modes. 
When the hierarchy is large, a low energy effective field theory (EFT) exists in which only curvature modes participate in the dynamics of perturbations.  
In this EFT heavy fields continue to have a significant role in the low energy dynamics, as their interaction with curvature modes reduces their speed of sound whenever the multi-field trajectory is subject to a {\it sharp turn} in target space.
Here we analyze under which general conditions this EFT remains a reliable description for the linear evolution of curvature modes. 
We find that the main condition consists on demanding that the rate of change of the turn's angular velocity stays suppressed with respect to the masses of heavy modes.
This {\it adiabaticity} condition allows the EFT to accurately describe a large variety of situations in which the multi-field trajectory is subject to {\it sharp turns}.
To test this, we analyze several models with turns and show that, indeed, the power spectra obtained for both the original two-field theory and its single-field EFT are identical when the adiabaticity condition is satisfied.
In particular, when turns are sharp and sudden, they are found to generate large features in the power spectrum, accurately reproduced by the EFT.

\end{abstract}

\end{titlepage}

\newpage


\section{Introduction}
\setcounter{equation}{0}

Cosmic inflation~\cite{Guth:1980zm} persists as the undisputed mechanism explaining the origin of primordial curvature perturbations~\cite{Mukhanov:1981xt} necessary to account for the Cosmic Microwave Background (CMB) anisotropies~\cite{Komatsu:2010fb, Larson:2010gs, Hlozek:2011pc} and the large scale structure of our universe~\cite{Sanchez:2005pi, Tegmark:2006az, Ho:2012vy, Seo:2012xy}. 
The fact that inflation is formulated within a field theoretical framework~\cite{Lyth:1998xn, Bassett:2005xm} makes it particularly compelling to test our ideas about fundamental theories, such as supergravity and string theory, characterized for consistently incorporating the gravitational strength among their couplings. 
Because these theories generically predict the existence of a large number of degrees of freedom, the need of a period of inflation at early times is found to impose strong restrictions on their interactions.
In particular, if inflation happened at sufficiently high energies, curvature perturbations could have strongly interacted with other degrees of freedom, implying a variety of observable effects departing from those predicted in standard single-field slow-roll inflation~\cite{Linde:1981mu, Albrecht:1982wi, Maldacena:2002vr}, including features in the power spectrum of primordial inhomogeneities~\cite{Starobinsky:1992ts, Polarski:1992dq, Chung:1999ve, Adams:2001vc, Ashoorioon:2006wc, Gong:2005jr, Ashoorioon:2008qr, Romano:2008rr, Tye:2008ef, Tye:2009ff, Barnaby:2010ke, Achucarro:2010da, Chen:2011zf, Park:2012rh}, large primordial non-Gaussianities~\cite{Linde:1996gt, Bartolo:2001cw, Bernardeau:2002jy, Lyth:2002my, Alishahiha:2004eh, Bartolo:2004if, Chen:2006nt} and isocurvature perturbations~\cite{Linde:1985yf, Starobinsky:1986fxa, Polarski:1994rz, DiMarco:2002eb, Gordon:2000hv, GrootNibbelink:2000vx, GrootNibbelink:2001qt, Lalak:2007vi, Choi:2008et}. 
A detection of any of these signatures would therefore represent an extremely significant step towards elucidating the fundamental nature of physics taking place during the very early universe.

Despite of its simplicity, the construction of satisfactory models of inflation within supergravity and string theory is known to constitute a notoriously hard challenge. 
The vacuum expectation values (v.e.v.'s) of scalar fields participating of the inflationary dynamics must evolve along flat directions of the scalar potential's landscape for a sufficiently long time. 
But because the interaction strength of these theories is of a gravitational nature, the scalar potential is naturally subject to changes of order $1$ when the scalar fields v.e.v.'s traverse distances of the order of the Planck scale.
This translates into the well known $\eta$-problem of supergravity and string theory~\cite{Copeland:1994vg, McAllister:2005mq, Covi:2008cn, Hardeman:2010fh}, where second derivatives of the scalar potential $V''$ are typically of order equal or larger than $H^2$ (where $H$ is the universe's expansion rate) therefore impeding the slow-roll evolution of the fields.%
\footnote{Another major obstacle towards the construction of models of inflation within string theory is related to the stabilization of moduli. See for instance refs.~\cite{Covi:2008ea, Covi:2008zu} for a discussion on the stabilization of moduli in supergravity and string theory.} 
However, this problem may be cured if the theory contains a set of shift symmetries at non-perturbative level, ensuring the existence of exactly flat directions in the potential~\cite{Stewart:1994ts}. 
Then, if these symmetries are mildly broken, the expected result is an inflationary scenario with a large mass hierarchy between the modular fields representing flat directions and the rest of the scalar fields, expected to have masses much larger than $H$~\cite{Conlon:2005jm, Grimm:2007hs, Linde:2007jn, Silverstein:2008sg, Baumann:2009ni}.

\begin{figure}[t!]
\begin{center}
\includegraphics[scale=0.38]{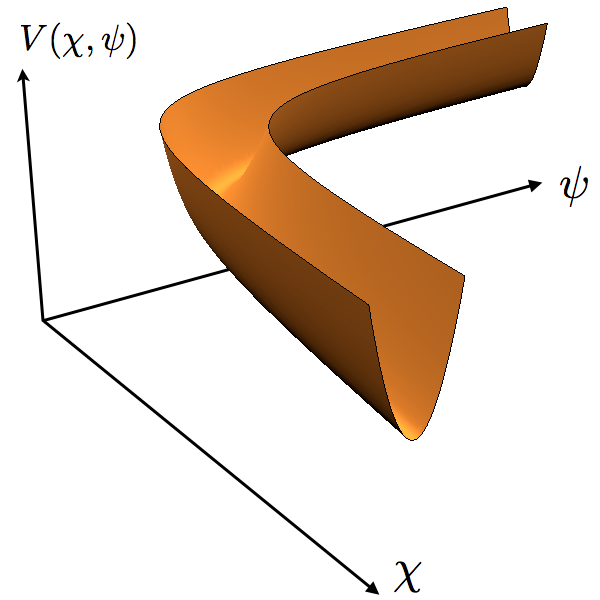}
\caption{\footnotesize The figure illustrates a prototype example of a multi-field potential (depending on two fields $\chi$ and $\psi$) with a mass hierarchy in which the flat direction is subject to a turn.}
\label{turn-prototype}
\end{center}
\end{figure}

Conventional wisdom dictates that UV-degrees of freedom with masses $M \gg H$ necessarily have a marginal role in the low energy dynamics of curvature modes. 
After these heavy degrees of freedom are integrated out, one expects a low energy effective field theory (EFT) for curvature perturbations where UV-physics is parametrized by nontrivial operators suppressed by factors of order $H^2/M^2$.
The resulting low energy EFT is therefore expected to offer negligible departures from a truncated version of the same theory, wherein heavy fields are simply disregarded from the very beginning.
However, general field theoretical arguments due allow for large sizable corrections to the low energy EFT~\cite{Cheung:2007st, Weinberg:2008hq}.
In the specific case of multi field models, there are special circumstances where the background inflationary dynamic is such that the interchange of kinetic energy between curvature perturbations and heavy degrees of freedom may be dramatically enhanced~\cite{Chen:2009we, Tolley:2009fg, Chen:2009zp, Achucarro:2010jv, Cremonini:2010sv, Cremonini:2010ua, Achucarro:2010da, Baumann:2011su, Shiu:2011qw, Achucarro:2012sm}.
For example, if the inflationary trajectory is subject to a sharp turn in such a way that the heavy scalar fields stay normal to the trajectory (see Figure~\ref{turn-prototype} for an illustration) then unsuppressed interactions ---kinematically coupling curvature perturbations with heavy fields--- are unavoidably turned on. 
As a consequence, if the rate of turn is large compared to the rate of expansion $H$, the impact of heavy physics on the low energy dynamics becomes substantially amplified, introducing large non-trivial departures from a naively truncated version of the theory. 

In the particular case of two field models ---at linear order in the fluctuations--- heavy fields are identified with isocurvature perturbations, and their role is reduced to modify the speed of sound $c_s$ of curvature perturbations at the effective field theory level.
The result is a non-trivial effective single-field theory where the time dependence of $c_s$ is dictated by the specific shape of the two-field background trajectory, in such a way that departures from unity  $c_s \neq 1$ exist whenever the trajectory is subject to a turn. 
More specifically, one finds that the speed of sound depends on the angular velocity $\dot \theta$ characterizing the turn as
\be
c_s^{-2} = 1 + 4 \dot \theta^2 / M_{\rm eff}^2,
\ee
where $M_{\rm eff}^2 = M^2 -  \dot H \,  \mathbb{R} - \dot \theta^2$ is the effective mass of isocurvature perturbations, with $M$ the tree-level bare mass of heavy modes, and $\mathbb{R}$ the Ricci scalar of the scalar field target space. 
In this way, sudden turns of the trajectory translate into sudden time variations of $c_s$ (hence modifying the value of the sound horizon $c_s / H$) and therefore  generating features in the power spectrum of primordial inhomogeneities~\cite{Achucarro:2010da}.\footnote{For a recent discussion on features in the power spectrum generated by variations of the speed of sound see ref.~\cite{Park:2012rh}.}
Moreover, if the turn is such that $c_s \ll 1$, cubic interactions become unsuppressed~\cite{Achucarro:2012sm}, implying large levels of primordial non-Gaussianities in the distribution of curvature perturbations~\cite{Chen:2006nt}.

The purpose of this article is to study two-field models of inflation characterized by a large mass hierarchy.\footnote{For other interesting work regarding non-trivial effects on the dynamics of curvature perturbations coming from massive degrees of freedom, see for instance refs.~\cite{Rubin:2001in, Jackson:2010cw, Jackson:2011qg}.}
We are particularly interested in assessing the general conditions under which the EFT deduced by integrating out the heavy field remains a reliable  description of the inflationary dynamics. 
We show that the main condition simply consists on the requirement that the rate of variation of the angular velocity $\dot \theta$ characterizing the turn stays suppressed with respect to the effective mass $M_{\rm eff}$ of heavy modes. That is:
\be
\left| \frac{d}{d t} \ln \dot \theta  \right| \ll M_{\rm eff} . \label{intro:adiabatic-cond}
\ee
We show that this {\it adiabaticity} condition is sufficiently mild, as it still allows for the effective field theory to describe, with great accuracy, a large variety of situations where very sharp turns take place ({\it i.e.} situations where $|\dot \theta | \gtrsim H$).
We check this condition by studying various models with turns and compare the power spectra of these models obtained from both, the full two-field inflationary model and the respective effective field theory.
We find that turns are able to generate large features in the power spectra, with the amplitude of these features depending on how large departures of $c_s$ from unity are.

This work is organized as follows: In Section~\ref{sec:review} we provide a self contained review on two-field models of inflation and summarize the main known results concerning the existence of  mass hierarchies. 
Then, in Section~\ref{sec:effective-theory} we offer a simple derivation of the general class of single-field EFT emerging from two-field models with mass hierarchies, and derive condition (\ref{intro:adiabatic-cond}) dictating the validity of this theory in terms of background quantities.
In Section~\ref{sec:sudden-turns} we discuss the different classes of turns and deduce the type of reactions that turns have on the background inflationary trajectory.
In particular, we study the case of sudden turns,  where the inflationary trajectory is subject to a single turn for a brief period of time $\Delta t$ smaller (or much smaller) than an $e$-fold $(\Delta t \lesssim H^{-1})$. 
Then, in Section~\ref{sec:examples},  we consider two toy models  and compute the power spectrum for different cases of turns. 
There we show that, consistent with (\ref{intro:adiabatic-cond}), the effective field theory remains reliable as long as $\Delta t \gg 1 / M_{\rm eff}$, where $M_{\rm eff}$ is the effective mass of the heavy field. 
We also show that large features on the power spectrum are easily produced, with the details of the effects depending on the different parameters characterizing the type of turns. 
Finally, in Section~\ref{sec:conclusions} we offer our concluding remarks to this work.


\section{Two-field inflation} \label{sec:review}
\setcounter{equation}{0}

In this section we summarize the main results coming from previous work related to the study of multi-field inflation~\cite{GrootNibbelink:2000vx, GrootNibbelink:2001qt, Achucarro:2010da}.\footnote{For other general approaches to multi-field models of inflation, see refs.~\cite{Langlois:2008mn, Gao:2009qy, Peterson:2010np, Gong:2011uw}.}  We shall specialize these results to the particular case of two-field models, to be studied in detail throughout this work. To start with, let us consider a non-canonical scalar field system with an action given by
\begin{equation}
S=\int d^4 x\sqrt{-g}\left[\frac{1}{2} R-\frac{1}{2}g^{\mu\nu} \gamma_{a b}\partial_{\mu}\phi^a\partial_{\nu}\phi^b-V(\phi)\right] , \label{total-action}
\end{equation}
where $R$ is the Ricci scalar constructed out of the spacetime metric $g_{\mu\nu}$ (notice that we are working in units where the Planck mass is set to unity $M_{\rm Pl}=1$). Additionally, $V(\phi)$ is the scalar field potential  and $\gamma_{a b}$ with $a = 1 , 2$ is the sigma model metric describing the abstract geometry of the scalar space spanned by the pair of fields $\phi^{1}$ and $\phi^{2}$. It is extremely useful to adopt a covariant notation with respect to the geometrical space offered by the scalar fields. This will allow us to deduce general results without making any reference to particular models in which $\gamma_{a b}$ and $V(\phi)$ acquire specific dependences on the fields. We therefore define a set of Christoffel symbols given by 
\be
\Gamma^{a}_{b c} = \frac{1}{2} \gamma^{a d} (\partial_b \gamma_{d c}  + \partial_c \gamma_{b d} - \partial_d \gamma_{b c} ) ,
\ee
where $\gamma^{a b}$ is the inverse sigma model metric. Then, the equations of motion for the scalar fields are found to be
\be
\Box \phi^a + \Gamma^{a}_{b c} \partial_\mu \phi^b \partial^{\mu} \phi^{c} - V^a = 0,
\ee
where  $ V^a \equiv \gamma^{a b} V_b$ with $V_{b} = \partial_b V$. We will also encounter the need of introducing the Ricci scalar, defined as $\mathbb{R} = \gamma^{a b} \mathbb{R}^{c}{}_{a c b}$, where $\mathbb{R}^{a}{}_{b c d}$ is the Riemann tensor given by:
\be
\mathbb{R}^{a}{}_{b c d} = \partial_{c} \Gamma^{a}_{b d} - \partial_{d} \Gamma^{a}_{b c} + \Gamma^{a}_{c e} \Gamma^{e}_{d b} -  \Gamma^{a}_{d e} \Gamma^{e}_{c b} .
\ee
Because we are specializing our analysis to two-field models, the Riemann tensor depends on a single degree of freedom, and therefore may be expressed in terms of the Ricci scalar $\mathbb{R}$ as:
\be
\mathbb{R}_{a b c d} = \frac{1}{2} \mathbb{R} (\gamma_{a c} \gamma_{b d} - \gamma_{a d} \gamma_{c b} ).
\ee
In what follows we proceed to study the dynamics of this system by considering separately the homogeneous and isotropic background and the perturbations of the system.

\subsection{Homogeneous and isotropic backgrounds}

\begin{figure}[t!]
\begin{center}
\includegraphics[scale=0.5]{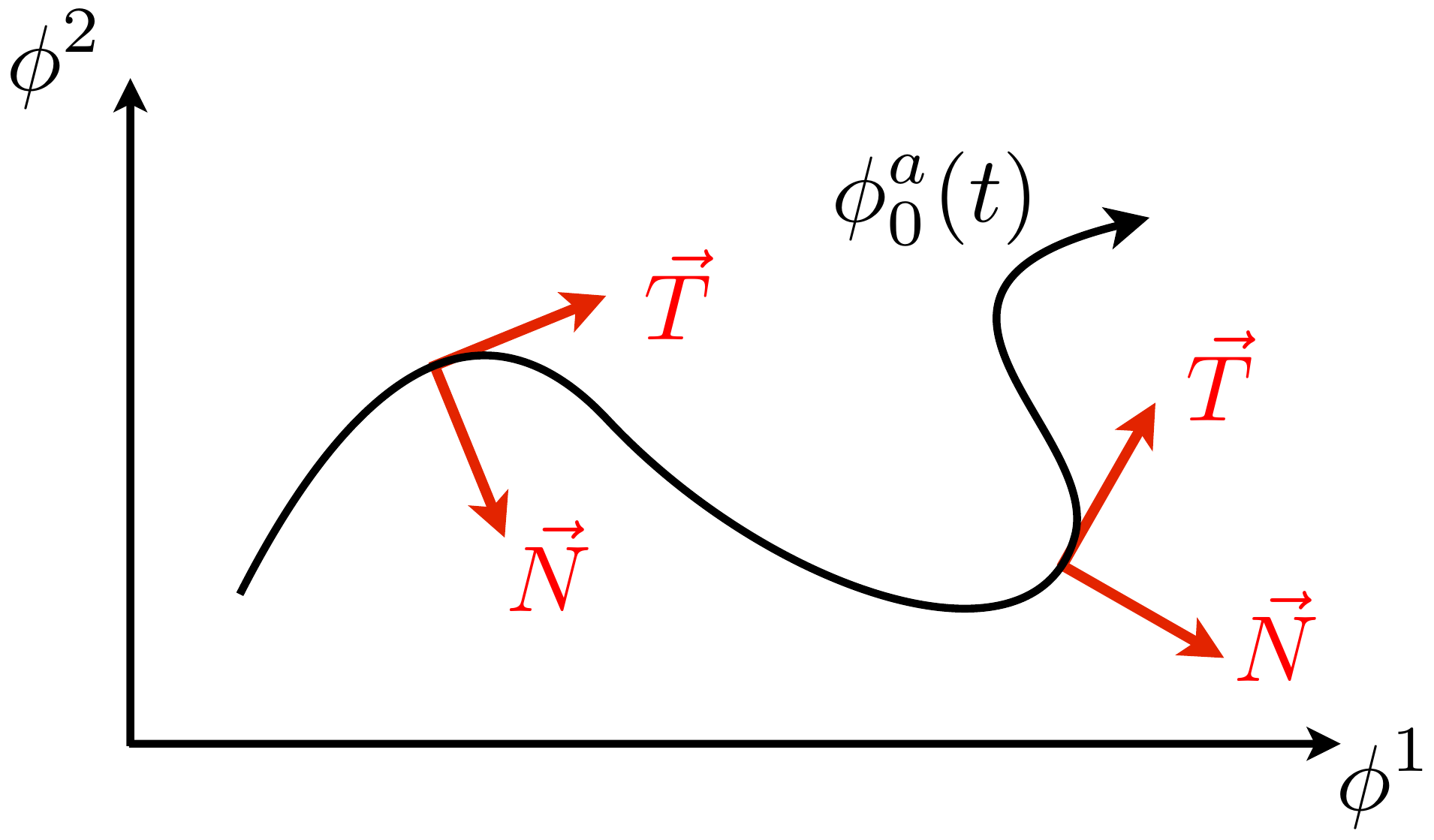}
\caption{\footnotesize Relative orientation of the vector fields $T^a$ and $N^a$ defined with respect to the background solution $\phi_0^a(t)$.}
\label{fig-T-and-N}
\end{center}
\end{figure}

Let us first devote our attention to homogeneous and isotropic cosmological backgrounds characterized by a scalar field solution $\phi^a = \phi^a_0(t)$ only dependent on time. For this we consider a flat Friedmann-Robertson-Walker metric of the form
\begin{equation}
ds^2=-dt^2+a^2(t)\delta_{ij}dx^idx^j ,
\end{equation}
where $a(t)$ is the scale factor describing the expansion of flat spatial foliations. Then, the equations of motion determining the evolution of the system of fields $a(t)$, $\phi_0^1(t)$ and $\phi_0^2(t)$ are given by
\bea
\frac{D}{dt}\dot{\phi}_0^a+3H\dot{\phi}^a_0+V^a=0 , \label{scalar-field-homogeneous-eqs} \\
3 H^2 = \frac{1}{2}\dot{\phi_0^2}+V , \label{Friedmann-eq}
\eea
where $H= \dot{a} / a$ is the rate of expansion. Equation (\ref{scalar-field-homogeneous-eqs}) corresponds to the equation of motion derived by varying the action with respect to $\phi^a$. There, we have introduced a covariant time derivative ${D}_{t}$ defined to satisfy
\be
\frac{D}{dt} X^a = \dot X^a + \Gamma^{a}_{b c} \dot \phi_0^b X^c ,
\ee
where $X^a = X^a(t)$ is an arbitrary vector field with the property of  transforming like $X^{a'} = \frac{\partial \phi^{a'}}{\partial \phi^{b}} X^b$ under a general field reparametrizations $\phi^{a'} = \phi^{a'}(\phi^1 , \phi^2)$. On the other hand, eq.~(\ref{Friedmann-eq}) (Friedmann's equation) determines the expansion rate in terms of the energy density of the system $\rho = \frac{1}{2}{\dot \phi_0^2}+V $. There, we are using the following notation to define the scalar field velocity $\dot \phi_0$:
\be
\dot \phi_0^2 \equiv \gamma_{a b} \dot \phi_0^a \dot \phi_0^b . 
\ee
By combining  (\ref{scalar-field-homogeneous-eqs}) and (\ref{Friedmann-eq})  we may derive the following useful relation:
\be
\dot{H} =-\frac{\dot \phi_0^2}{2 }  .
\ee
Given a set of initial conditions for the scalar fields, there will exist a unique solution $\phi_0^a(t) = (\phi_0^1(t) , \phi_0^2(t))$ defining a curve in field space parametrized by cosmic time $t$. To characterize this curve, it is convenient to construct a set of orthogonal unit vectors $T^a$ and $N^a$ in such a way that, at a given time $t$, $T^a(t)$ is tangent to the path, and $N^a(t)$ is normal to it. We may define this set of vectors as:\footnote{We use the metric $\gamma_{ab}$ and its inverse $\gamma^{ab}$ to lower and rise indices whenever it is required. In particular, we have $T_{a} = \gamma_{ab}T^b$ and $N^{a} = \gamma^{ab} N_b$.}
\bea
T^a &=& \dot \phi_0^a / \dot \phi_0 , \\
N_a &=& \left( \det  \gamma \right)^{1/2} \epsilon_{a b} T^{b} ,
\eea
where $\epsilon_{a b}$ is the two dimensional Levi-Civita symbol with $\epsilon_{11} = \epsilon_{22} = 0$ and $\epsilon_{12} = - \epsilon_{21} = 1$. These definitions ensure that $T_a T^a = N_a N^a  = 1$ and $T^a N_a = 0$. Notice that $N_a$ has a fixed orientation with respect to the path, as shown in Figure~\ref{fig-T-and-N}. 
These two unit vectors may be used to project the scalar field equations of motion in (\ref{scalar-field-homogeneous-eqs}) along the two orthogonal directions. Projecting along $T^a$, one finds:
\be
\ddot{\phi}_0+3H\dot{\phi_0}+V_{\phi}=0,
\ee
where $V_{\phi} \equiv \nabla_\phi V$, with $\nabla_{\phi} \equiv T^a \partial_a$. On the other hand, projecting along $N^a$, one obtains the relation
\be
\frac{DT^a}{dt}=-\frac{V_N}{\dot{\phi_0}}N^a, \label{time-deriv-T}
\ee
where $V_N = N^a \partial_a V$. It is customary to define dimensionless parameters accounting for the time variation of various background quantities. These are the so called slow-roll parameters $\epsilon$, $\eta^a$ and we define them as:
\be
\epsilon  \equiv  -\frac{\dot{H}}{H^2} , \qquad  \eta^a  \equiv  -\frac{1}{H\dot{\phi_0}}\frac{D\dot{\phi_0^a}}{dt}.
\ee
Notice that $\eta^a$ is a two dimensional vector field telling us how fast $\dot \phi_0^a$ is changing in time. We may decompose $\eta^a$ along the normal and tangent directions by introducing two independent parameters $\eta_{||}$ and $\eta_{\perp}$ as
\be
\eta^a =\eta_{\parallel}T^a+\eta_{\bot}N^a.
\ee
Then, one finds that 
\bea
\eta_{\parallel} = -\frac{\ddot{\phi_0}}{H\dot{\phi_0}} , \\
\eta_{\bot}  =  \frac{V_N}{\dot{\phi_0}H} . \label{def-eta-perp}
\eea
Notice that $\eta_{\parallel} $ may be recognized as the usual $\eta$ slow-roll parameter in single field inflation. On the other hand $\eta_{\bot}$ tells us how fast $T^a$ rotates in time, and therefore it parametrizes the rate of turn of the trajectory followed by the scalar field dynamics. This may be seen more clearly by using (\ref{time-deriv-T}) together with (\ref{def-eta-perp}) to deduce the following relations:
\bea
\frac{DT^a}{dt} &=& - H \eta_{\perp} N^a , \\
\frac{DN^a}{dt} &=& + H \eta_{\perp} T^a .
\eea
Thus, if $\eta_{\perp} = 0$, the vectors $T^a$ and $N^a$ remain constant along the path. On the other hand, if  $\eta_{\perp} > 0$, the path turns to the left, whereas if $\eta_{\perp} < 0$ the turn is towards the right..  
The value of $\eta_{\perp}$ is therefore telling us how quickly the angle determining the orientation of $T^a$ is varying in time. By calling this angle $\theta$ we may therefore do the identification
\be
\dot \theta \equiv  H \eta_{\perp} . \label{angle-variation}
\ee
With the help of this definition,  one deduces that the ratio of curvature $\kappa$ characterizing the turning trajectory, is given by 
\be
\kappa^{-1} \equiv  | \dot \theta | / \dot \phi_0 . \label{kappa-def}
\ee  
As in conventional single-field inflation, the background dynamics may be understood in terms of the values of the dimensionless parameters $\epsilon$, $\eta_{||}$ and $\eta_{\perp}$.  For instance, slow roll inflation will happen as long as: 
\be
\epsilon \ll 1 , \qquad |\eta_{||}| \ll 1 . \label{slow-roll-regime}
\ee 
These two conditions ensure that both $H$ and $\dot \phi_0$ evolve slowly during the period of interest. On the other hand, it is interesting to notice that a large variation of $\eta_{\perp}$  does not necessarily imply a violation of the slow-roll regime (\ref{slow-roll-regime}). We will analyze this statement in full detail in Section~\ref{sec:sudden-turns} where we study the effect of sharp sudden turns on the dynamics of this class of system.

\subsection{Perturbation Theory}

\begin{figure}[t!]
\begin{center}
\includegraphics[scale=0.54]{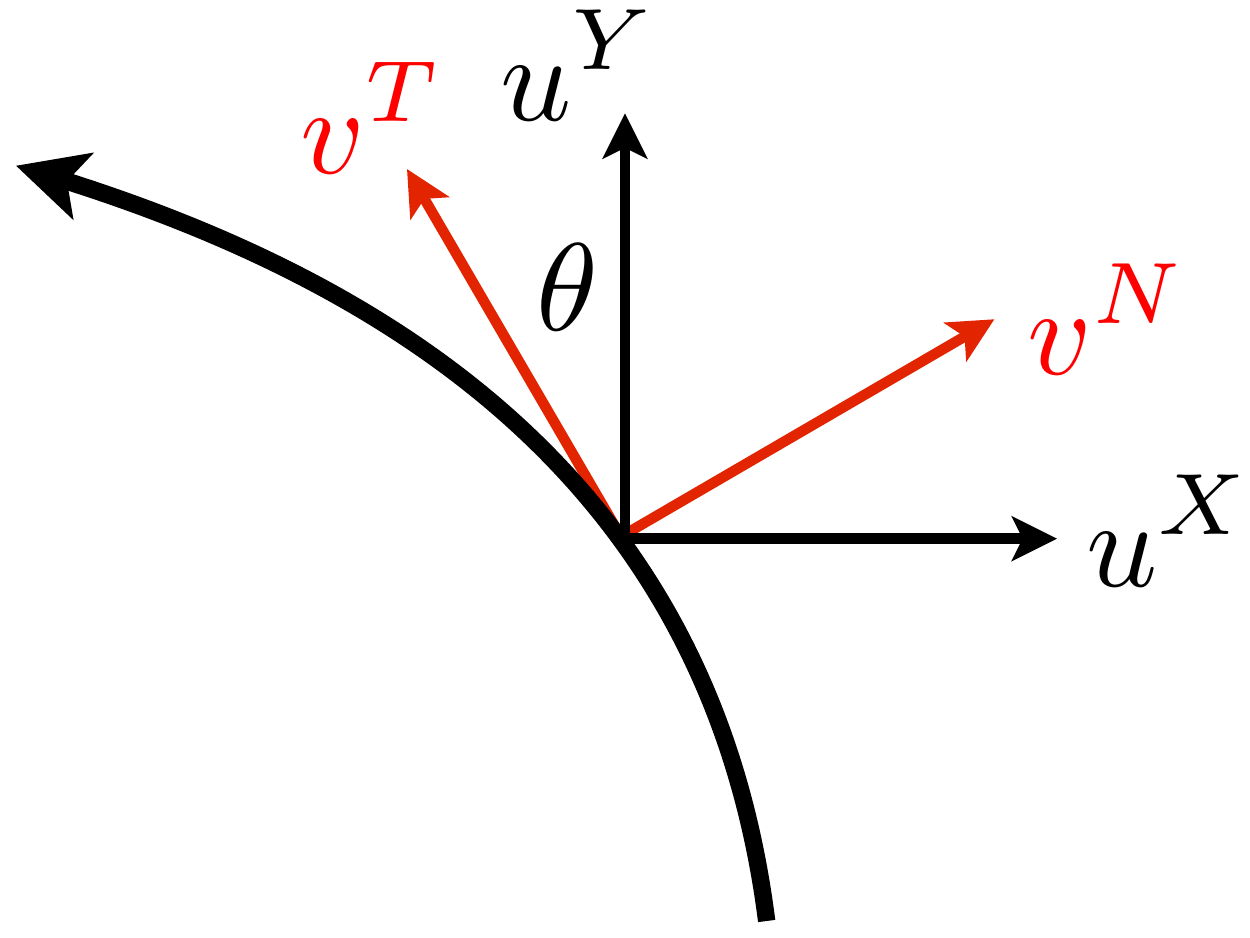}
\caption{\footnotesize The $u$-fields represent fluctuations with respect to a fixed local frame, whereas the $v$-fields represent fluctuations with respect to the path (parallel and normal).}
\label{fig-v-u}
\end{center}
\end{figure}

We now consider the dynamics of scalar perturbations parametrizing departures from the homogeneous and isotropic background $a(t)$ and $\phi_0^a(t)$. This may be done by defining perturbations $\delta\phi^a(t,\textbf{x})$ as: 
\be
\phi^a(t,\textbf{x})=\phi_0^a(t)+\delta\phi^a(t,\textbf{x}).
\ee 
Instead of directly working with $\delta\phi^a(t,\textbf{x})$, it is more convenient to work with gauge invariant fields $v^{T}$ and $v^{N}$ given by\footnote{Notice that these fields are projections of the form $v^T = a T_a Q^a$ and $v^T = a T_a Q^a$ where the $Q^a$ fields are the usual Mukhanov-Sasaki variables $Q^a\equiv\delta\phi^a+\frac{\dot{\phi}^a}{H}\psi$~\cite{Sasaki:1986hm, Mukhanov:1988jd}.}
\bea
v^{T} &=&  a \, T_a \delta\phi^a+ a \frac{\dot \phi}{H}\psi , \\
v^{N} &=&  a \, N_a  \delta\phi^a ,
\eea
where $\psi$ is the scalar perturbation of the spatial part of the metric (proportional to $\delta_{ij}$) in flat gauge. It is useful to consider a second set of fields $(u^X , u^Y)$ in addition to $(v^T, v^N)$. Let us consider the following time dependent rotation in field space
\be
\left(\begin{array}{c} u^{X} \\ u^{Y}\end{array}\right) \equiv  R (\tau)  \left(\begin{array}{c} v^{N} \\ v^{T} \end{array}\right),
\ee
where the time dependent rotation matrix $R (\tau)$ is defined as
\be
 R (\tau) = \left(\begin{array}{cc} \cos \theta(\tau) & - \sin \theta(\tau)  \\  \sin \theta(\tau)  & \cos \theta(\tau)  \end{array}\right) , \qquad \theta(\tau) = \theta_0 + \int^{\tau}_{-\infty} \!\!  d \tau \,   a H \eta_{\perp} ,
\ee
where $\theta_0$ is the value of $\theta (\tau)$ at $\tau \to - \infty$.
The rotation angle $\theta(\tau)$ precisely accounts for the total angle covered by all the turns during the inflationary history up to time $\tau$, and coincides with the definition introduced in eq.~(\ref{angle-variation}). Figure \ref{fig-v-u} illustrates the relation between the $v$-fields introduced earlier and the canonical $u$-fields.
To continue, the equations of motion for the canonically normalized fields are
\begin{equation}
\frac{d^2u^I}{d\tau^2}-\nabla^2u^I+\left[R (\tau)\Omega R^t (\tau)\right]^I_{\ J}u^J=0, \qquad I = X,Y ,
\end{equation}
where $R^t $ represents the transpose of $R$. In addition, $\Omega$ is the mass matrix for the $v$-fields, with elements given by
\bea
\Omega_{TT}&=&-a^2 H^2(2+2\epsilon-3 \eta_{\parallel}+\eta_{\parallel} \xi_{\parallel}-4\epsilon \eta_{\parallel}+2 \epsilon^2 - \eta_{\perp}^2 ),\\
\Omega_{NN}&=&-a^2 H^2(2-\epsilon)+a^2 (V_{N N} + H^2  \epsilon\mathbb{R} ),\\
\Omega_{TN}&=&a^2 H^2\eta_{\bot}(3+\epsilon-2\eta_{\parallel}  - \xi_{\bot}) ,
\eea
where $ \xi_{\parallel} =  - \dot \eta_{||} / (H \eta_{||})$ and $\xi_{\bot} = - \dot \eta_{\perp} / (H \eta_{\perp})$. Additionally, we have defined the tree level mass $V_{NN}$ as the second derivative of the potential projected along the perpendicular direction $V_{NN} = N^a N^b \nabla_a \nabla_b V$. To finish, expanding the original action (\ref{total-action}) to quadratic order in terms of the $u$-fields, one finds:
\begin{equation}
S=\frac{1}{2}\int d\tau d^3x \left\lbrace\sum_I \left(\frac{ d u^I}{d\tau}\right)^2-(\nabla u^I)^2-[R(\tau)\Omega R^t(\tau)]^I_{\ J} u^I u^J \right\rbrace.  \label{canonical-action}
\end{equation}
Thus, we see that the fields $u^I = (u^X, u^Y)$ correspond to the canonically normalized fields in the usual sense. Given that these fields are canonically normalized, it is now straightforward to impose Bunch-Davies conditions on the initial state of the perturbations.

\subsection{Curvature and isocurvature modes}

Another useful field parametrization for the perturbations is in terms of curvature and isocurvature fields $\mathcal R$ and $\mathcal S$~\cite{Gordon:2000hv}. In terms of the $v$-fields, these are defined as:
\bea
\mathcal{R} &=& \frac{H}{a \dot \phi} v^{T}  , \\
\mathcal{S} &=& \frac{H}{a \dot \phi} v^{N} .
\eea
Instead of working directly with $\mathcal S$, it is in fact more convenient to define:
\be
\mathcal F = \frac{\dot \phi_0}{H} \mathcal S.
\ee
Then, the quadratic action for the pair $\mathcal R$ and $\mathcal F$ is found to be
\bea
S_{\rm tot} =  \frac{1}{2} \int \! d^4 x \, a^3 \bigg[  \frac{\dot \phi_0^2}{H^2} \dot {\mathcal R}  ^2  - \frac{\dot \phi_0^2}{H^2} \frac{(\nabla \mathcal R)^2}{a^2}  +   \dot {\mathcal F}  ^2  
- \frac{ (\nabla \mathcal F)^2 }{a^2}
+  4 \dot \phi_0 \eta_{\perp} \dot{\mathcal R} \mathcal F    - M_{\rm eff}^2 \mathcal F^2   \bigg] ,
 \quad \label{total-action-R-F}
\eea
where we have defined the effective mass $M_{\rm eff}$ of the heavy field  $\mathcal F$ as
\be
M_{\rm eff}^2 = V_{N N} + H^2  \epsilon\mathbb{R} - \dot \theta^2 ,  \label{effective-mass}
\ee
(recall that $\dot \theta =  H \eta_{\perp}$). It may be noticed that the reason behind the appearance of the term $- \dot \theta^2$ in $M_{\rm eff}^2$ is due to the fact that the potential receives a correction coming from the centripetal force experimented by the turn. This introduces a centrifugal barrier to the effective potential felt by the heavy modes.
The equations of motion for this system of fields is then
\bea
\ddot {\mathcal R} +  (3  + 2 \epsilon - 2 \eta_{||}) H \dot  {\mathcal R} - \frac{\nabla^2  {\mathcal R} }{a^2} &=& - 2 \frac{H^2}{\dot \phi_0} \eta_{\perp} \left[ \dot {\mathcal F} + (3 - \eta_{||} - \xi_{\perp}) H {\mathcal F} \right] , \\
\ddot {\mathcal F} +  3  H \dot  {\mathcal F} - \frac{\nabla^2  {\mathcal F} }{a^2} + M_{\rm eff}^2  {\mathcal F} &=& 2 \dot \phi_0 \eta_{\perp} \dot {\mathcal R} . \label{eq-of-motion-F}
\eea
Notice that the configuration $\mathcal R =$~constant and $\mathcal F = 0$ constitutes a non trivial solution to the system of equations. Since $\mathcal F$ is assumed to be heavy, the configuration $\mathcal F = 0$ is reached shortly after horizon exit, and the curvature mode $\mathcal R$ will necessarily become frozen. For this reason, in the presence of mass hierarchies, we may only concern ourselves with curvature perturbations and disregard isocurvature components after inflation.

\subsection{Power spectrum}

From the observational point of view, the main quantities of interest coming from inflation are its predicted $n$-point correlation functions characterizing fluctuations. These quantities provide all the relevant information about the expected distribution of primordial inhomogeneities that seeded the observed CMB anisotropies. It is of particular interest to compute two-point correlation functions, corresponding to the variance of inhomogeneities' distribution. To deduce such quantities we have to consider the quantization of the system, and this may be achieved by expanding the canonical pair $u^X$ and $u^Y$ in terms of creation and annihilation operators $a^{\dag}_{\alpha}({\bf k})$ and  $a_{\alpha}({\bf k}) $ respectively, as
\be
u^I(\tau,x) =   \int \frac{d^3k}{(2 \pi)^{3/2}} \sum_{\alpha} \left[ e^{i {\bf k} \cdot {\bf x} } u^{I }_{\alpha}({\bf k} , \tau) a_{\alpha}({\bf k})  + e^{- i {\bf k} \cdot {\bf x} } u^{I *}_{\alpha}({\bf k} , \tau) a^{\dag}_{\alpha}({\bf k})  \right] ,
\ee
where $\alpha = 1,2$ labels the two modes to be encountered by solving the second order differential equations for the fields $u^{I }_{\alpha}({\bf k} , \tau)$. In order to satisfy the conventional field commutation relations, the mode solutions need to satisfy the additional constraints  consistent with the equations of motion:\footnote{See refs.~\cite{GrootNibbelink:2001qt} and~\cite{Achucarro:2010da} for a more detailed discussion of the quantization of these type of system. }
\be
\sum_{\alpha} \left(  u_{\alpha}^{I} \frac{u_{\alpha}^{J *}}{d \tau} -   u_{\alpha}^{I *} \frac{u_{\alpha}^{J}}{d \tau}   \right) = i \delta^{I J}.
\ee
 By examining the action (\ref{canonical-action}) one sees that in the short wavelength limit $k^2 / a^2 \gg \Omega$, where $\Omega$ symbolizes both eigenvalues of the matrix $\Omega$, the equation of motion for the $u$-fields reduce to 
\be
\frac{d^2u^I}{d\tau^2}-\nabla^2u^I =0, \qquad I = X,Y , \label{short-wavelength-limit}
\ee
allowing us to choose the following initial conditions for the fields
\be
u^X_{\alpha} ({\bf k} , \tau)  = \frac{e^{- i k \tau}}{\sqrt{2 k}}  \delta_{\alpha}^{1},   \qquad   u^Y ({\bf k} , \tau)  = \frac{e^{- i k \tau}}{\sqrt{ 2 k}}    \delta_{\alpha}^{2} . \label{initial-state}
\ee
Notice that here we have chosen to associate the initial state $\alpha = 1$ with the field direction $X$ and $\alpha = 2$ with the field direction $Y$. This identification is in fact completely arbitrary and does not affect the computation of two-point correlation functions. In other words, we could modify the initial state (\ref{initial-state}) by considering an arbitrary (time independent) rotation on the right hand side, without changing the prediction of observables.

Then, given the set of solutions $u_{\alpha}^{X}$ and $u_{\alpha}^Y$, one finds that the two-point correlation function associated to curvature modes $\mathcal R$ is given by:
\bea
\mathcal{P}_{\mathcal R}(k,\tau) &=& \frac{k^3 }{2\pi^2}\sum_{\alpha} {\mathcal R}_{\alpha}(k,\tau) {\mathcal R}_{\alpha}^{*}(k,\tau) ,  \label{two-point-corr-definition} 
\eea
where $\mathcal R_{\alpha}$ is related to the pair $u_{\alpha}^{X}$ and $u_{\alpha}^Y$ by the field redefinitions described in the previous sections. When (\ref{two-point-corr-definition})  is evaluated at the end of inflation, for wavelengths $k$ far away from the horizon ($k/a \ll H$), it corresponds to the power spectrum of curvature modes.  For completeness, let us mention that one may also define the two-point correlation function $\mathcal{P}_{\mathcal S}(k,\tau)$ and the cross-correlation function $\mathcal{P}_{\mathcal R  \mathcal S}(k,\tau)$ in analogous ways. But, as previously stated, because of the assumed mass hierarchy these contributions may be completely disregarded.


\section{Effective Field Theory} \label{sec:effective-theory}
\setcounter{equation}{0}

It is possible to deduce an effective theory for the curvature mode $\mathcal R$ by integrating out the heavy field $\mathcal F$ when $M_{\rm eff}^2 \gg H^2$, provided that certain additional conditions are met. To see this, let us first briefly analyze the expected evolution of the fields $\mathcal R$ and $\mathcal F$ when the trajectory is turning at a constant rate ($\dot \theta = $ constant). To start with, because we are dealing with a coupled system of equations for $\mathcal R$ and $\mathcal F$, in general we expect the general solutions for $\mathcal R$ and $\mathcal F$ to be of the form~\cite{Achucarro:2010jv}
\bea
\mathcal R & \sim & \mathcal R_{+} e^{- i \omega_{+} t} + \mathcal R_{-} e^{- i \omega_{-} t} , \label{frequency-R} \\
\mathcal F & \sim & \mathcal F_{+} e^{- i \omega_{+} t} + \mathcal F_{-} e^{- i \omega_{-} t} , \label{frequency-F}
\eea
where $\omega_{+}$ and $\omega_{-}$ denote the two frequencies at which the modes oscillate. 
The values of $\omega_{+}$ and $\omega_{-}$ will depend on the mode's wave number $k$ in the following way: In the regime $k / a \gg M_{\rm eff}$, both fields are massless and therefore oscillate with frequencies of order $\sim k/a$. As the wavelength enters the intermediate regime $M_{\rm eff} \gg k / a \gg H$ the degeneracy of the modes break down and the frequencies become of order 
\be
\omega_{-} \sim k/a,  \qquad \omega_{+} \sim M_{\rm eff}.
\ee
Subsequently, when the modes enter the regime $k/a < H$ the contributions coming from $\omega_+$ will quickly decay and the contributions coming from $\omega_{-}$ will freeze (since they are massless).

Notice that the amplitudes $\mathcal R_{+}$ and $\mathcal F_{-}$ necessarily arise from the couplings mixing curvature and isocurvature perturbations, and therefore they vanish in the case $\eta_{\perp} =  \dot \theta / H = 0$.  Additionally, on general grounds, the amplitudes $\mathcal F_{+} $ and  $\mathcal R_{+} $ are expected to be parametrically suppressed by  $k / M_{\rm eff}$ in the regime $M_{\rm eff} \gg k / a$, and therefore we may disregard high frequency contributions to (\ref{frequency-R}) and (\ref{frequency-F}). Then, in the regime $M_{\rm eff} \gg k / a$, time derivatives for $\mathcal F$ can be safely ignored in the equation of motion (\ref{eq-of-motion-F}) and we may write:
\be
- \frac{\nabla^2  {\mathcal F} }{a^2} + M_{\rm eff}^2  {\mathcal F} = 2 \dot \phi_0 \eta_{\perp} \dot {\mathcal R} . \label{eq-of-motion-F-2}
\ee
(Because $H \ll M_{\rm eff}$, we may also disregarded the friction term $3 H \dot {\mathcal F}$). This leads to an algebraic relation between $\mathcal F$ and $\dot {\mathcal R}$ in Fourier space given by:
\bea
 {\mathcal F}_{\mathcal R} &=& \frac{2 \dot \phi_0 \eta_{\perp} \dot {\mathcal R}}{\left( k^2 / a^2 + M_{\rm eff}^2 \right)} , \label{F-algebraic-R}
\eea
which precisely tells us the dependence of low frequency contributions $\mathcal F_{-}$ in terms $\mathcal R_{-} $ defined in eqs.~(\ref{frequency-R}) and (\ref{frequency-F}).
To continue, we notice that (\ref{eq-of-motion-F-2}) is equivalent to disregard the term $\dot {\mathcal F}^2$ of the kinetic term in the action  (\ref{total-action-R-F}). Keeping this in mind, we can replace (\ref{F-algebraic-R}) back into the action and obtain an effective action for the curvature perturbation $\mathcal R$, given by\footnote{This way of integrating heavy modes has also been employed to deduce an effective theory for the linear propagation of gravitons in bigravity theories~\cite{Atal:2011te}, where a massive graviton kinetically interacts with the massless one. When the massive graviton is integrated out, one is left with a massless graviton with a non-trivial speed of propagation that depends on the background.}
\bea
S_{\rm eff} =  \frac{1}{2} \int \! d^4 x \, a^3 \frac{\dot \phi_0^2}{H^2} \bigg[ \frac{ \dot {\mathcal R}  ^2 }{c_s^2(k)}  
- \frac{ k^2 \mathcal R^2 }{a^2}   \bigg] ,
 \quad \label{total-action-R-effective}
\eea
where $c_s$ is the speed of sound of adiabatic perturbations, given by:
\be
c_s^{-2}(k) = 1 + \frac{4 H^2 \eta_{\perp}^2 }{k^2 / a^2 + M_{\rm eff}^2 } . \label{eq:speed-of-sound}
\ee
In deriving this expression we have assumed that $\dot \theta$ remained constant. In the more general case where $\dot \theta$ is time dependent we expect transients that could take the system away from the simple behavior shown in eqs.~(\ref{frequency-R}) and (\ref{frequency-F}), and the effective field theory could become invalid. The validity of the effective theory will depend on whether the kinetic terms for $\mathcal F$ in eq.~(\ref{eq-of-motion-F}) can be ignored, and this implies the following condition on $\mathcal F_{\mathcal R}$ of eq.~(\ref{F-algebraic-R}):
\be
| \ddot {\mathcal F}_{\mathcal R}   | \ll  M_{\rm eff}^2 | \mathcal F_{\mathcal R} | . \label{F-vs-M}
\ee
Now, recall that unless there are large time variations of background quantities, the frequency of $\mathcal R$ is of order $\omega_-  \sim k / a$. Thus, any violation of condition (\ref{F-vs-M}) will be due to the evolution of background quantities, which will be posteriorly transmitted to $\mathcal R$. This allows us to ignore higher derivatives of $\dot {\mathcal R}$ in (\ref{F-vs-M}) and simply rewrite it in terms of background quantities as:
\be
\left|   \frac{d^2}{d t^2} \left( \frac{2 \dot \phi_0 \eta_{\perp}}{\left( k^2 / a^2 + M_{\rm eff}^2 \right)}  \right)   \right| \ll M_{\rm eff}^2 \left| \frac{2 \dot \phi_0 \eta_{\perp}}{\left( k^2 / a^2 + M_{\rm eff}^2 \right)} \right| . \label{first-adiabatic-cond}
\ee
This relation expresses the adiabaticity condition that each mode $k$ needs to satisfy in order for the effective field theory to stay reliable. To further simplify this relation, we may take into consideration the following points: (1) When $k^2 / a^2 \gg M_{\rm eff}^2$ the two modes decouple (recall eq.~(\ref{short-wavelength-limit})) and the turn has no influence on the evolution of curvature modes. On the other hand, in the regime $k^2 / a^2 \lesssim M_{\rm eff}^2$, contributions coming from the time variation of $k^2 / a^2$ are always suppressed compared to $M_{\rm eff}^2$ due to the fact that we are assuming $H^2 \ll M_{\rm eff}^2$.  (2) We observe that the main background quantity parametrizing the rate at which the turn happens is $\eta_{\perp} =  \dot \theta / H$. Quantities such as $\dot \phi_0$ and $H$, which describe the evolution of the background along the trajectory, will therefore only be affected by the turn through the time dependence of $\eta_{\perp}$. This implies that time derivatives of these quantities will be less sensitive to the turn than $\eta_{\perp}$ itself, and therefore their contribution to (\ref{first-adiabatic-cond}) will necessarily be subsidiary.\footnote{Here we are implicitly assuming that {\it parallel} quantities such as $\dot \phi_0$ and $H$ will not have variations larger than $\eta_{\perp}$ due to other effects, unrelated to the turns (such as steps in the potential).} (3) Similarly, the rate of change of $M_{\rm eff}^2$ will necessarily be at most of the same order than $\dot \theta$. 
Then, by neglecting time derivatives coming from $\dot \phi_0$,  $H$, $k^2/a^2$ and $M_{\rm eff}$ and focussing on the order of magnitude of the various quantitates appearing in (\ref{first-adiabatic-cond}) we can write instead a simpler expression given by:
\be
 \left|  \frac{d^2}{d t^2}   \dot \theta  \right|   \ll  M_{\rm eff}^2  \left| \dot \theta \right| . \label{second-adiabatic-cond}
\ee
Actually, a simpler alternative expression may be obtained by conveniently reducing the number of time derivatives, and disregarding effects coming from the change in sign of $\dot \theta$:
\be
\left| \frac{d}{d t} \ln \dot \theta \right| \ll M_{\rm eff} . \label{eq-var-c_s-M_eff}
\ee
This adiabaticity condition simply states that the rate of change of the turn's angular velocity must stays suppressed with respect to the masses of heavy modes, which otherwise would become excited. Notice that, we may also choose to express this relation in terms of the variation of the speed of sound, which is a more natural quantity from the point of view of the effective field theory:
\be
\left| \frac{d}{d t} \ln (c_s^{-2} - 1) \right| \ll M_{\rm eff} .
\ee

To finish, we can estimate the order of magnitude of departures between the full solution for $\mathcal R$ and the one appearing in the EFT. For this, let us write $\mathcal F = \mathcal F_{\mathcal R} + \delta \mathcal F$ where $\mathcal F_{\mathcal R}$ is the adiabatic result of (\ref{F-algebraic-R}) and $\delta \mathcal F$ denotes a departure from this value. Then $\delta \mathcal F$ respects the following equation of motion:
\be
\ddot { \delta \mathcal F } + 3 H \dot { \delta \mathcal F } + \left( \frac{k^2}{a^2} + M_{\rm eff} \right) \delta \mathcal F = - ( \ddot {\mathcal F_{\mathcal R} } +  3 H \dot{ \mathcal F_{\mathcal R} }).
\ee
Given that we are assuming that the behavior of the system is dominated by low frequency modes, we can consistently disregard the kinetic term $\ddot { \delta \mathcal F } + 3 H \dot { \delta \mathcal F }$ at the left hand side of this equation, but we cannot disregard the term $- ( \ddot {\mathcal F_{\mathcal R} } +  3 H \dot{ \mathcal F_{\mathcal R} })$ at the right hand side, which constitutes a source for $\delta {\mathcal F}$. Then, we deduce that 
\be
 \delta \mathcal F = -  \frac{ \ddot {\mathcal F_{\mathcal R} } +  3 H \dot{ \mathcal F_{\mathcal R} }}{ k^2 / a^2 + M_{\rm eff} } .
\ee
Then, we may compute the derivatives appearing in the right hand side by using the effective equation of motion (coming from the variation of the effective action  (\ref{total-action-R-effective})) to express $\ddot{\mathcal R}$ in terms of $\dot {\mathcal R}$ whenever it becomes necessary. We obtain
\be
\mathcal F = \mathcal F_{\mathcal R} \left[ 1 + \mathcal O \left( \frac{ ( \ddot \theta / \dot \theta)^2 }{k^2/a^2 + M_{\rm eff}^2} \right) +  \mathcal O \left( \frac{\delta_{||} H^2 }{k^2/a^2 + M_{\rm eff}^2} \right)   \right] , \label{F-FR-error}
\ee
where $\delta_{||}$ represents terms of order $\epsilon$ and $\eta_{||}$. Finally, eq.~(\ref{F-FR-error}) allows us to deduce that the EFT expressed in (\ref{total-action-R-effective}) is only accurate up to operators of the form:
\be
S_{\rm tot} = S_{\rm eff} +  \frac{1}{2} \int \! d^4 x \, a^3 \,  (c_s^{-2} - 1) \, \dot {\mathcal R}^2   \left[ \mathcal O \left( \frac{ ( \ddot \theta / \dot \theta)^2 }{ M_{\rm eff}^2} \right) +  \mathcal O \left( \frac{\delta_{||} H^2 }{ M_{\rm eff}^2} \right)   \right].
\ee
This result expresses the validity of the effective field theory (\ref{total-action-R-effective}), and shows with eloquence the order of magnitude of the expected discrepancy with the exact two field solution for $\mathcal R$. In the following section we verify that indeed the adiabatic condition (\ref{eq-var-c_s-M_eff}) constitutes a good guide to discriminate the validity of the effective theory (\ref{total-action-R-effective}). 


\section{Turning trajectories}  \label{sec:sudden-turns}
\setcounter{equation}{0}

\begin{figure}[b!]
\begin{center}
\includegraphics[scale=0.48]{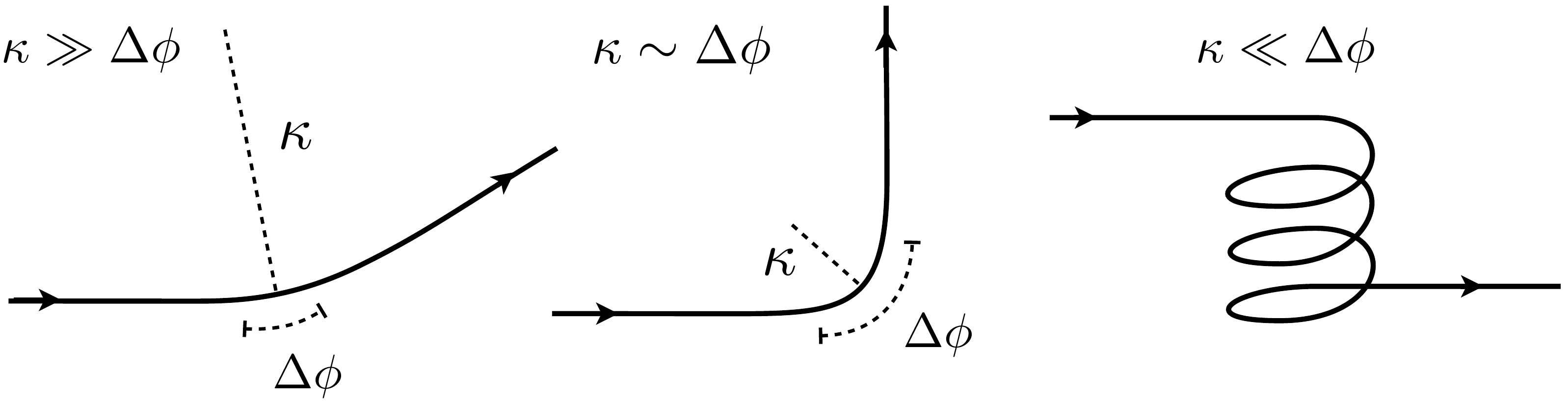}
\caption{\footnotesize Examples of turns according to the relative size of $\Delta \phi$ and $\kappa$. In the case $\kappa \ll \Delta \phi$ the target space requires a non trivial topology.}
\label{various-turns}
\end{center}
\end{figure}

We now study the consequences of turning trajectories on the dynamics of fluctuations. We start this analysis by considering the particular case of sudden turns, where the potential $V(\phi)$ and/or the sigma model metric $\gamma_{ab}$ entering the action (\ref{total-action}) are such that the inflationary trajectory becomes non-geodesic for a brief period of time, smaller than an $e$-fold. In order to characterize this class of turns it is useful to introduce the arc distance $\Delta \phi$ covered by the scalar field's v.e.v. in target space, when the turn takes place. Given the radius of curvature $\kappa$ characterizing a turn, defined in eq.~(\ref{kappa-def}), we may define $\Delta \phi$ through the relation
\be
\Delta \phi \equiv   \kappa | \Delta \theta |,  \label{Delta-theta-Delta-phi}
\ee
where $\Delta \theta = \int H \eta_{\perp} dt$ corresponds to the total angle covered during a sudden turn. It is clear that in two-field canonical models $\Delta \phi$ can be at most of order $\kappa$ ($\Delta \phi \lesssim \kappa$), and that in order to have turns with $\Delta \phi \gg \kappa$ one needs a non-canonical model with a scalar geometry with a topology allowing for such situations. Figure~\ref{various-turns} shows various examples of turns according to the total angle $\Delta \theta$ covered by a turn (which is determined by the relative size of $\Delta \phi$ and $\kappa$).

Notice that the arc length $\Delta \phi$ implies a timescale $T_\perp$ characterizing the overall duration of a turn. This is simply given by:
\be
T_\perp \equiv \frac{\Delta \phi}{\dot \phi}  . \label{def-time-Delta}
\ee
Then, because  $\eta_{\perp} = \dot \theta /H \simeq  \Delta \theta /(T_\perp H) $, we can use eq.~(\ref{Delta-theta-Delta-phi}) to derive the following estimation for the value of $\eta_{\perp}$ characterizing a particular turn:
\be
\eta_{\perp} \sim  \frac{1}{ H T_\perp} \frac{\Delta \phi}{\kappa}. \label{eta-perp-in-terms-of-T}
\ee
Notice that a turn requires that the inflationary trajectory departs from the flat minima of the potential (otherwise $V_{N} = 0$ and eq.~(\ref{def-eta-perp}) would require $\eta_\perp = 0$). Then, because the turn happens suddenly during a brief period of time, the various interactions present in the theory will inevitably make the background trajectory oscillate about the flat locus of the potential, with a period $T_{M}$ given by
\be
T_{M} \equiv \frac{1}{M_{\rm eff}} . \label{def-time-M}
\ee
In other words, a turn cannot be arbitrarily sharp without waking up these background fluctuations. Recall that we are interested in models where $M_{\rm eff} \gg H$, and therefore we necessarily have $T_{M} \ll H^{-1}$. If present, it is clear that such oscillations will dominate the behavior of the trajectory whenever $T_M$ is of the same order or larger than $T_\perp$ ($T_M \gtrsim T_\perp$). In fact, the adiabaticity condition (\ref{eq-var-c_s-M_eff}) precisely translates into the following condition involving these two timescales
\be
T_{\perp} \gg T_{M}, \label{adiabatic-cond-times}
\ee
which is consistent with the notion that the effective field theory will remain valid as long as heavy fluctuations are not participating of the low energy dynamics.

\subsection{Displacement from the flat minima}

Given certain turn (characterized by $\Delta \phi$ and $\kappa$) we can estimate the perpendicular displacement of the trajectory away from the flat minima of the potential while the turn takes place. By calling this quantity $\Delta h$, we can roughly relate it to other background quantities through the relation $V_N \simeq M_{\rm eff}^2 \Delta h$. Then, inserting this result back into eq.~(\ref{def-eta-perp}) we obtain
\be
 \frac{\Delta h}{\kappa} \simeq  \frac{\Delta \phi^2 }{   \kappa^2} \frac{T_M^2}{T_\perp^2} , \label{eq-displacement}
\ee
where we made use of eqs.~(\ref{def-time-Delta}) and (\ref{def-time-M}). This relation gives us the relative size of background oscillations $\Delta h$ compared to the radius of curvature $\kappa$ of the turn. We see that the size of the displacement depends on the total angle covered by the turn $\Delta \theta = \Delta \phi / \kappa$ and the ratio $T_M / T_{\perp}$ between the two relevant timescales. In what follows we briefly analyze the two relevant regimes posed by these two timescales.

\subsection{Adiabatic turns $T_M \ll T_\perp$} \label{sec:case-TM-less-than-Tperp}

\begin{figure}[h!]
\begin{center}
\includegraphics[scale=0.5]{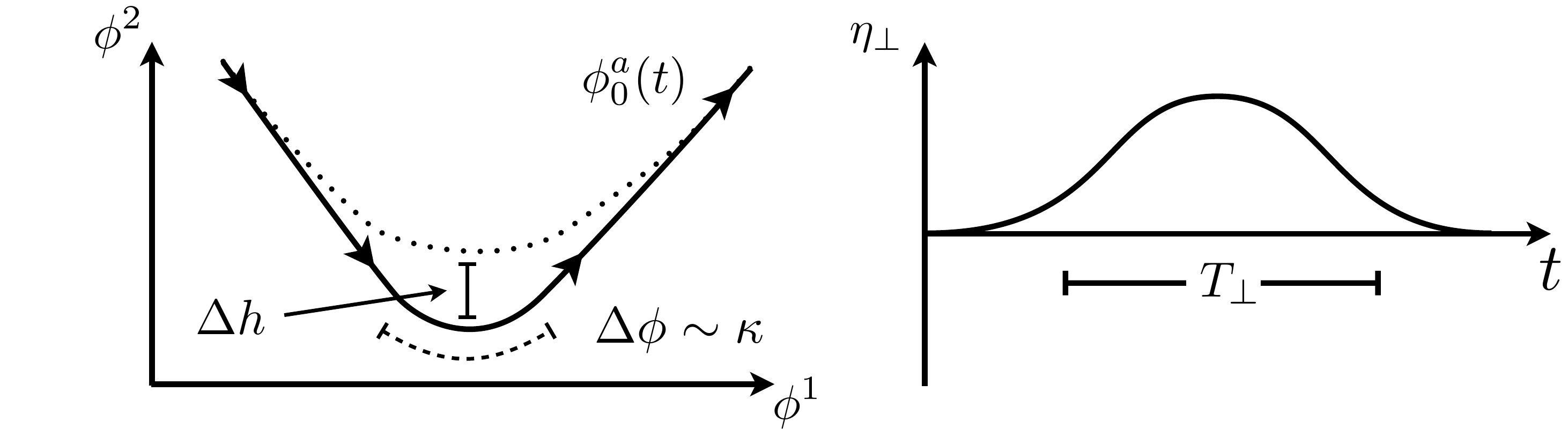}
\caption{\footnotesize The figure illustrates the case of a turn for which the adiabatic condition $T_{M}  \ll T_{\perp}$ is respected. In this case the turn happens adiabaticaly, in the sense that the timescale $T_M$ plays no role during the turn. If $\Delta h / \kappa \sim 1$ the displacement from the flat minima is large and the speed of sound will be reduced considerably.}
\label{plot-adiabatic-turns}
\end{center}
\end{figure}

If the turn is such that $T_M \ll T_\perp$, then the adiabatic condition (\ref{eq-var-c_s-M_eff}) is satisfied and the system admits an effective field theory of the form (\ref{total-action-R-effective}) as deduced in the previous section. This means that we can parametrize effects in terms of a reduced speed of sound $c_s$, which synthesizes all the nontrivial information regarding the heavy physics. Putting together eqs.~(\ref{eq:speed-of-sound}) and (\ref{eta-perp-in-terms-of-T}) we see that the speed of sound is given by
\be
c_s^{-2}  \simeq 1 +  \frac{\Delta \phi^2 }{  \kappa^2  } \frac{T_M^2}{T_\perp^2}.
\ee
Given that the effective theory requires $T_{M}  \ll T_{\perp}$, the only way of having large non-trivial departures from conventional single field inflation is by having a large ratio $\Delta \phi / \kappa \gg 1$.
This implies that the total angle $\Delta \theta = \Delta \phi / \kappa$ covered by the turn must extend for several cycles, and the only way of achieving this consistently is by considering models with non-trivial sigma model metrics. In particular, to produce sizable changes of the speed of sound (say of order one) we require:
\be
 \frac{\Delta \phi^2 }{   \kappa^2} \gtrsim  \frac{T_\perp^2}{T_M^2} .
\ee
Comparing with (\ref{eq-displacement}), we see that this is equivalent to have a large displacement from the flat minima of the potential. However, because the timescale $T_M$ is much smaller than $T_{\perp}$, the displacement happens adiabatically, and background fluctuations are not turned on. Correspondingly, the dimensionless parameter $\eta_\perp$ is a smooth function of time with a characteristic timescale given by $T_\perp$.  Figure~\ref{plot-adiabatic-turns} illustrates this situation.

\subsection{Non-adiabatic turns  $T_M \gtrsim T_\perp$}

\begin{figure}[h!]
\begin{center}
\includegraphics[scale=0.5]{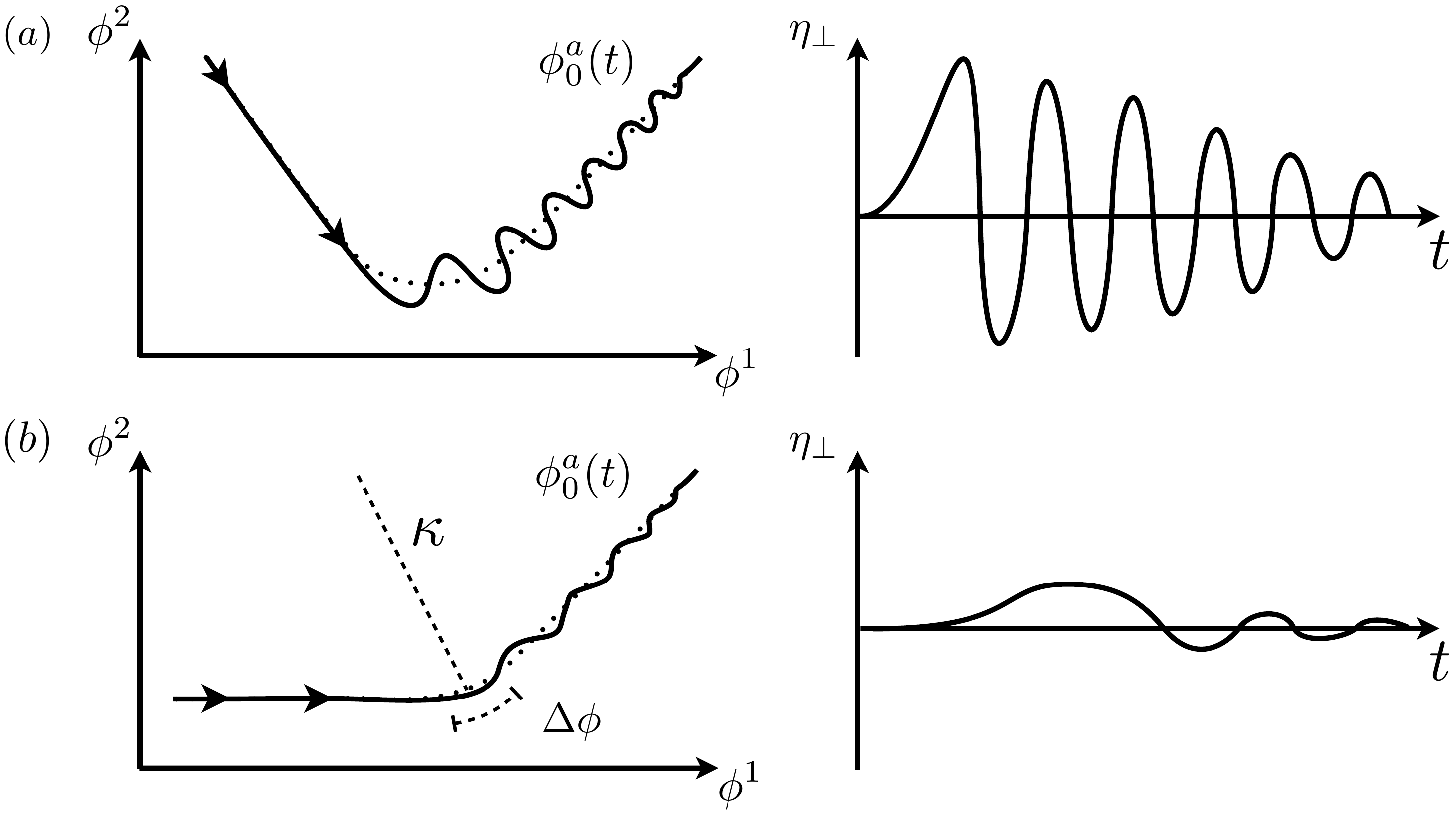}
\caption{\footnotesize Examples of non-adiabatic turns characterized by $T_{M}  \ll T_{\perp}$. Case (a)  shows a typical example in which $\Delta \phi_0 \sim \kappa$, whereas case (b) shows a situation for which $\Delta \phi_0 \ll \kappa$.}
\label{plot-non-adiabatic-turns}
\end{center}
\end{figure}

In this case the trajectory moves away from the flat minima of the potential quickly, and the background trajectory will inevitably start oscillating about this locus. The amplitude of these oscillations will be given by $\Delta h$ as in eq.~(\ref{eq-displacement}). If $\Delta h \gtrsim \kappa$ then the oscillations are large and they completely dominate the behavior of the background trajectory. Needless to say, the effective field theory would offer a poor representation of the evolution of curvature perturbations, and the original two-field theory would be needed to study the system. Figure~\ref{plot-non-adiabatic-turns}-(a) shows this type of situation for the case $T_\perp \sim T_M$. There, the trajectory is subject to an initial kick that lasts $T_M$, and after that continues oscillating at a period given by $T_M$. Consequently, heavy oscillations will overtake the background trajectory, which translates into a heavily oscillating $\eta_{\perp}$ as a function of time. This means that instead of a single turn we end up with a succession of turns. However, after these transient turns have taken place, the overall angle $\Delta \theta = \int H \eta_{\perp} dt$ will correspond to the turn determined by the flat minima of the potential.

On the other hand, if $\Delta h \ll \kappa$, then the amplitude of the oscillations are small. In this case $\Delta \phi / \kappa  \ll   T_{\perp} / T_M$, implying, after considering eq.~(\ref{eta-perp-in-terms-of-T}), that the angular velocity is small compared to the mass of heavy modes:
\be
 H |\eta_{\bot}| =  |\dot \theta|  \ll M_{\rm eff} . \label{condition-sudden-turns-M}
\ee
In this case the impact of background oscillations on the evolution of perturbations is negligible. Despite of this fact, since we are in the regime $T_M \gtrsim T_\perp$, the effective field theory continues to offer a poor representation of these effect, no matter how tiny they are. Figure~\ref{plot-non-adiabatic-turns}-(b) illustrates this situation.
%


\section{Examples of models with turns} \label{sec:examples}
\setcounter{equation}{0}

We now study examples of models where turns play a relevant role on the evolution of perturbations.  We will first consider the case of canonical models ---in which turns are uniquely due to the shape of the potential--- and later consider the case of models where the turns are due to the specific shape of the sigma model metric.

\subsection{Model 1: Sudden turns in canonical models} \label{sec:canonical-model}

\begin{figure}[t!]
\begin{center}
\includegraphics[scale=0.44]{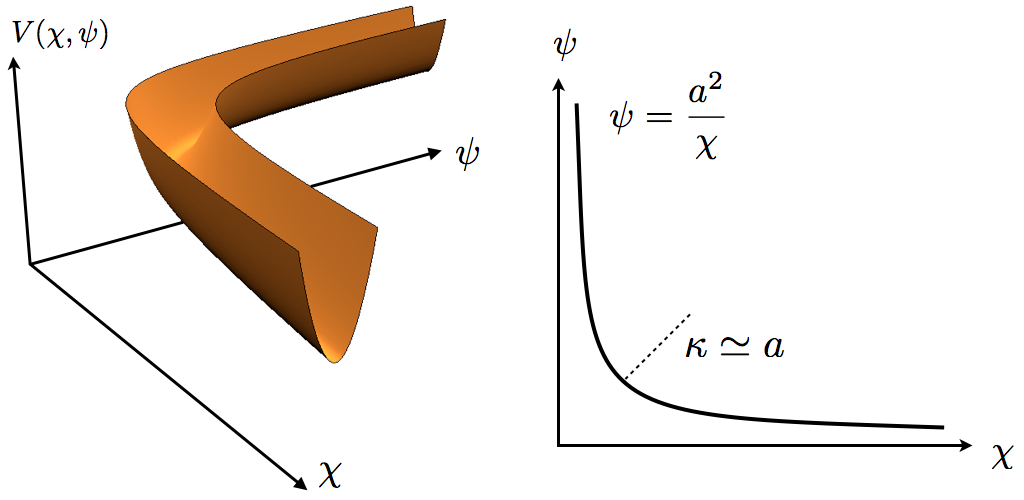}
\caption{\footnotesize The figure shows the section of the potential $V(\chi,\psi)$ of equation (\ref{example-potential}) containing the turn.}
\label{turn-potential}
\end{center}
\end{figure}

Let us start by considering the case of canonical models ($\gamma_{ab} = \delta_{ab}$) in which the turn is due solely to the shape of the potential. To simplify the present analysis, we only consider the relevant part of the potential where the turn takes place, and disregard any details concerning the end of inflation. Our aim is to understand what are the possible consequences of a turn on the generation of primordial inhomogeneities accessible to observations today. For this reason, we assume that the turn takes place precisely when presently accessible curvature perturbations were crossing the horizon, and choose cosmological parameters accordingly. Having said this, let us adopt the notation $\phi^a = ( \chi, \psi)$ and consider the following potential:
\be
V(\chi , \psi)=  V_0 + V_{\phi} ( \chi -  \psi ) + \frac{M^2}{2}  \frac{(\chi  \psi  - a^2 )^2}{ (\chi+\psi )^2} + \cdots . \label{example-potential}
\ee
In this expression, $V_0$ and $V_{\phi}$ are constants parametrizing the hight and slope of the flat direction in the potential. The third term in (\ref{example-potential}) has been added to ensure that a turn takes place. Notice that the third term vanishes for the locus of points defined by the equation:
\be
\psi =   a^2 / \chi  . \label{locus-turn-pot}
\ee
Away from this curve the slope of the potential becomes steep, with a quadratic growth characterized by the mass parameter $M$. 
Figure~\ref{turn-potential} shows the relevant part of the potential $V(\chi,\psi)$ containing the turn.
Because we are assuming a mass hierarchy, it may be anticipated that the inflationary trajectory will stay close to the curve defined by eq.~(\ref{locus-turn-pot}). Roughly speaking, the turn is located about the position $(\chi , \psi ) = (\sqrt{a} , \sqrt{a})$ and that its radius of curvature is of order $a$:
\be
\kappa \sim a.
\ee
The quantities $V_0$ and $V_{\phi}$ are chosen in such a way that the scalar fields v.e.v.'s approach the turn from the asymptotic direction $(\chi , \psi ) \to (0, + \infty)$. Once the turn is left behind, the v.e.v.'s. continue evolving towards the asymptotic direction $(\chi , \psi ) \to (+ \infty, 0)$, until inflation ends (see Figure~\ref{turn-potential}).

\subsubsection{Analysis of the model}

Before studying the exact evolution of the system with the help of numerics, let us estimate the behavior of the system by examining the parameters entering the potential.  First of all, if the potential is flat enough, the slow roll evolution of the fields imply that {\it away from the location of the turn} the following relations are satisfied:
\bea
3 H \dot \phi_0 \simeq V_{\phi} , \label{parallel-1} \\
3  H^2 \simeq V_0 .  \label{parallel-2} 
\eea
Notice that they imply that $\epsilon \simeq V_{\phi}^2 / 2 V_0^2 $. 
To continue, simple examination of the potential shows that the arc length $\Delta \phi$ characterizing the turn is of the same order than the radius of curvature $\kappa$ which, as stated, is of order $a$. Thus, we have $\kappa \simeq \Delta \phi \simeq a$. Then, putting together the previous expressions we find that time $T_{\perp}$ characterizing the duration of the turn, is given by
\be
T_{\perp} \simeq \frac{ a}{\sqrt{2 \epsilon} H}. \label{T-perp-turn-pot}
\ee
In terms of $e$-folds, the duration of the turn is given by
\be
\Delta N_{\perp} \simeq  \frac{ a}{\sqrt{2 \epsilon}} .  \label{Delta-N-def}
\ee
Then, using (\ref{eta-perp-in-terms-of-T}) we see that the value $\eta_{\perp}$ characterizing the turn is roughly given by:
\be
\eta_{\perp}  \simeq \frac{ \sqrt{2 \epsilon} }{a}. \label{eta-perp-in-terms-of-T-example-1}
\ee
With these relations at hand, we can now estimate the range of parameters for which the adiabatic condition~(\ref{eq-var-c_s-M_eff}) is satisfied. Using eq.~(\ref{adiabatic-cond-times}) with $T_M = 1 / M$, we deduce that the EFT will remain accurate as long as:
\be
\alpha \equiv \frac{T_\perp^2}{T_M^2} = \frac{M^2 a^2}{2 \epsilon H^2}  \gg  1 . \label{condition-example-pot}
\ee
Then, assuming that we are in the realms of validity of the theory, we find that the speed of sound is given by the following combination of parameters
\be
c_s^{-2}  \simeq 1 +  \frac{8 \epsilon H^2 }{ a^2 M^2   }, 
\ee
which, because of eq.~(\ref{condition-example-pot}), implies that $c_s \sim 1$, consistent with our general analysis of Section~\ref{sec:case-TM-less-than-Tperp}. 

\subsubsection{Numerics}

We now  present our numerical analysis of this model. In order to study this model, we have chosen the following fixed values (in units of Planck masses) for parameters associated to the flat direction of the potential:
\be
V_0 = 6.5 \times 10^{-9}  , \quad V_{\phi} = - 5.4 \times 10^{-11} .
\ee
Away from the turn, this choice of parameters ensure the following value for the rate of expansion $
H \sim 1.4 \times 10^{-5}$. Additionally, they imply values for the slow roll parameters given by 
\be
\epsilon \simeq - \eta \simeq 0.0033 ,
\ee
consistent with a power spectrum for curvature perturbations with a spectral index given by $n_s \simeq 0.98$ and an amplitude satisfying COBE's normalization. The other two parameters left are $a$ and $M$. These may be expressed in terms of $\Delta N_{\perp}$ and $\alpha$ introduced in eqs.~(\ref{Delta-N-def}) and (\ref{condition-example-pot}).

\begin{figure}[h!]
\begin{center}
\includegraphics[scale=0.75]{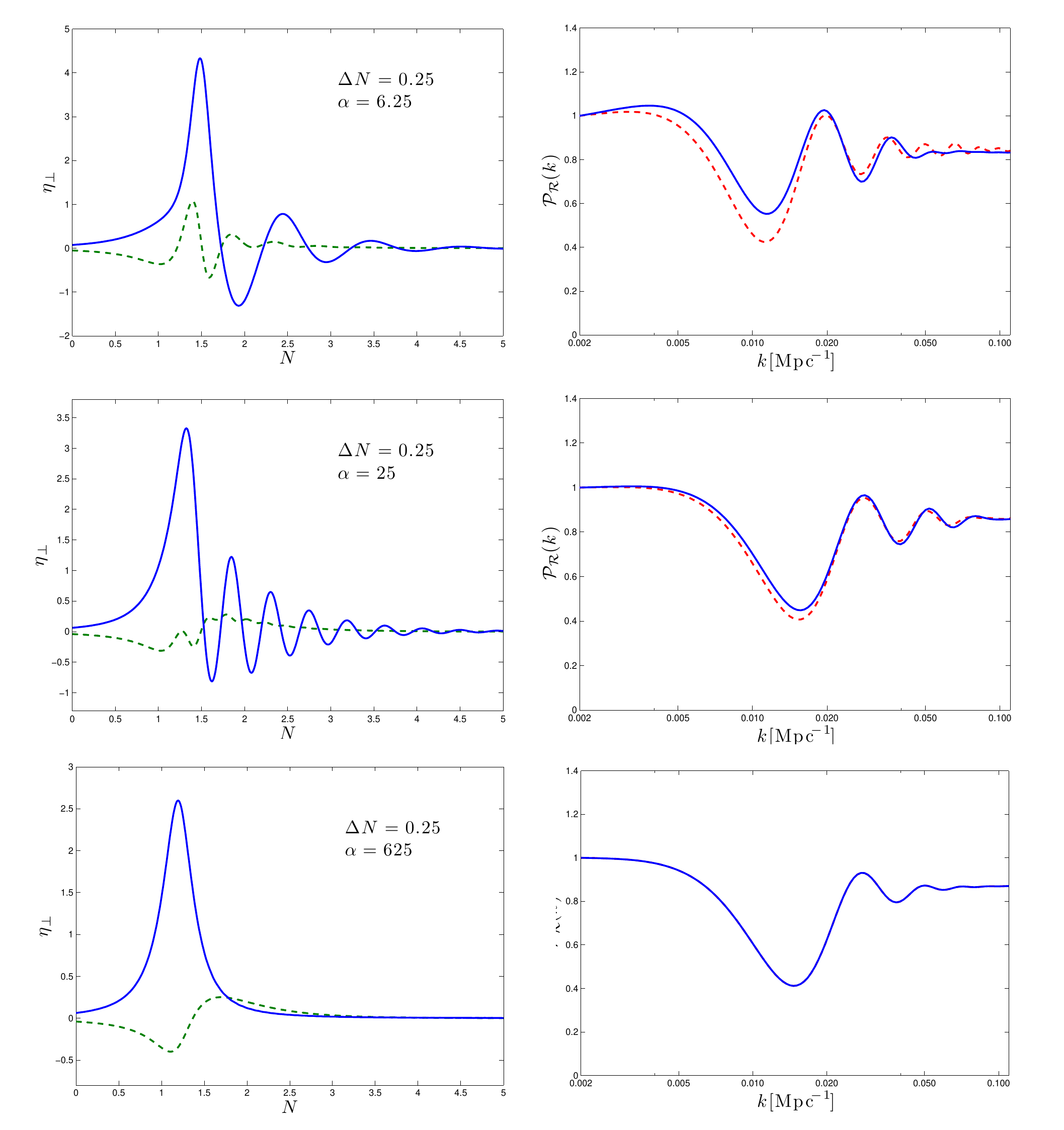}
\caption{\footnotesize  The figure shows $\eta_{\perp}$ and $\eta_{||}$ (left panels) and the resulting power spectra (right panels) for three choices of parameters for the potential of our model 1. From top to bottom: $\alpha = 6.25$, $25$ and $625$. In the case of the left panels, the blue solid line corresponds to $\eta_{\perp}$ whereas the green dashed line corresponds to $\eta_{||}$. In the case of right panels, the blue solid line corresponds to the power spectrum for the full two-field model, whereas the red dashed line corresponds to the power spectrum deduced using the EFT.
}
\label{examples-potential}
\end{center}
\end{figure}

Since we are interested in turns taking place during a period of time shorter than an $e$-fold ($T_{\perp} \lesssim H^{-1}$), we are required to choose $\Delta N_{\perp} \lesssim 1$.  Figure~\ref{examples-potential} shows the numerical results for the fixed value $\Delta N_{\perp} = 0.25$ and three choices for the mass $M$, given by $\alpha = 6.25$, $\alpha =25$ and $\alpha = 625$. On the left hand side of the figure we show the numerical solution for the $\eta_{\perp}$ and $\eta_{||}$ as functions of $e$-fold $N$. On the right hand side, we show the resulting power spectra for the relevant modes exiting the horizon when the turn takes place. For simplicity, we have normalized the power spectrum with respect to the amplitude determined by the largest scales (smallest values of $k$), which have been chosen to correspond to the largest scales characterizing horizon reentry today ($k_0 = 0.002$Mpc$^{-1}$). The blue solid line corresponds to the complete-two dimensional theory, whereas the red dashed line corresponds to the result predicted by the effective field theory. 

It may be seen that for $\alpha = 6.25$, the period of massive oscillations $T_{M}$ is of the same order than $T_{\perp}$, and consequently the rate of turn $\eta_{\perp}$ is dominated by oscillations (notice that these oscillations have a period of about $\sim 0.25$ $e$-folds, in agreement with our choice of parameters). This indicates that indeed the inflationary trajectory stays oscillating about the minimum once the turn has occurred (recall case (a) of Figure~\ref{plot-non-adiabatic-turns}). It may be also observed that $\eta_{||}$ is considerably affected by these oscillations, momentarily acquiring values as large as $\eta_{||} \sim 1$. This is however not enough to break down the overall slow-roll behavior of the background solution for two reasons: First, the value of $\epsilon$ does not change much during the turn (that is $\epsilon$ is found to be less sensitive to the turn than $\eta_{||}$). And second, after the turn takes place, the system goes back to small values of $\eta_{||}$ quickly. Given that the adiabatic condition is far from being satisfied ($\alpha$ is close to 1), it comes to no surprise that both power spectra differ considerably. 

The second case $\alpha =25$ shows an intermediate situation where $T_{M}$ is smaller than $T_{\perp}$ but still able to generate small oscillations about the minimum of the potential. In this case the adiabaticity condition is mildly violated, which is reflected on the small discrepancy between both power spectra. Finally, the case $\alpha = 625$ shows a situation where $T_{M}$ is much smaller than $T_{\perp}$. In this case heavy modes are not excited enough and the turn happens smoothly, which is reflected on the behavior of both  $\eta_{\perp}$ and $\eta_{||}$ as functions of $e$-folds.  In addition, we now see that both power spectra agree considerably.  
 It may call our attention the persistence of a large feature in the power spectra even for the case $\alpha = 625$ where both, versions of the theory agree. In this case the speed of sound $c_s \simeq 1$ during the turn and we infer that the feature cannot be due to the fast variation of the speed of sound. Instead, the feature is due to the large variation of parallel background quantities, specially $\eta_{||}$, as during the turn the trajectory is forced to go up and down. In the examples of the following model we will examine a rather different situation where the features in the power spectrum are due to large variations of the speed of sound $c_s$.

\subsection{Model 2: Sudden turns induced by the metric}

We now study a case in which the  turn is  due to the sigma model metric $\gamma_{a b}$. We will adopt the following notation $\phi^1 = \chi$ and $\phi^2 = \psi$, and consider the following separable potential
\be
V(\chi , \psi) = V_0 + V_{\phi} \, \chi  +  \frac{M_{\psi}^2}{2} \psi^2 .
\ee
Just like in the previous example, $V_0$ and $ V_{\phi}$ are constants parametrizing the flat part of the potential driving inflation. As before, we omit in our analysis any detail concerning the end of inflation.
For this potential, if the system were canonical ($\gamma_{a b} =  \delta_{a b}$), there would be no turns and the v.e.v. of the heavy field $\psi$ would stay sitting at its minimum $\psi_0 = 0$. To produce a single turn with the help of the metric, we consider the following model:
\be
\gamma_{a b} = \left(\begin{array}{cc} 1 & \Gamma (\chi) \\ \Gamma (\chi) & 1 + \Gamma^2(\chi) \end{array}\right),
\ee
where
\be
\Gamma(\chi) = \frac{\Gamma_{0} }{2 }  \left( 1 + \tanh \left[ 2 ( \chi - \chi_0)/\Delta \chi  \right]  \right)  .
\ee
Notice that $\Gamma(\chi)$ is a function that grows monotonically from the asymptotic value $\Gamma = 0$ at $\chi \to - \infty$ to the asymptotic value $\Gamma = \Gamma_0$ at $\chi \to + \infty$. The transition takes place at $\chi = \chi_0$ and is characterized by the width parameter $\Delta \chi$.  Notice that once $\Gamma$ reach the constant value $ \Gamma_0$, the metric becomes canonical again (which means that one can find a new parametrization of the fields in which $\gamma_{a b} =  \delta_{a b}$). We will consider values of $V_0$ and $V_\phi$ such that the field $\chi$ evolves from $\chi \to - \infty$ to $\chi \to + \infty$.

\subsubsection{Analysis of the model}

It is clear that $\Gamma(\chi)$ parametrizes departures from the canonical configuration $\gamma_{a b} =  \delta_{a b}$. The potential is such that it will force the trajectory to stay on the locus of points $\psi = 0$. However, since the geometry of the target space is non-trivial, the trajectory will be subject to a turn. To estimate the effects of the turn on the relevant background quantities, notice first that the timescale associated to heavy fluctuations about the minimum $\psi = 0$ is trivially given by:
\be
T_{M} = \frac{1}{M_{\psi}} .
\ee
Second, since the trajectory remains close to $\psi =0$ (due to the mass hierarchy), the arc length of the turn in target space $\Delta \phi$ will be equal to the width of the function $\Gamma(\chi)$:
\be
\Delta \phi = \Delta \chi.
\ee
Then, it immediately follows that the timescale $T_{\perp}$ characterizing the duration of the turn is simply given by:
\be
T_{\perp} =  \frac{ \Delta \chi}{\sqrt{2 \epsilon}},
\ee
where $\epsilon = V_{\phi}^2 / 2 V_0^2$. Again, we may express this period of time terms of $e$-folds, which is given by
\be
\Delta N_{\perp} \simeq  \frac{\Delta \chi}{\sqrt{2 \epsilon}} .  \label{Delta-N-def-2}
\ee
Then, in terms of the parameters of the model, the adiabaticity condition reads
\be
\alpha \equiv \frac{T_\perp^2}{T_M^2} = \frac{M_{\psi}^2 \Delta \chi^2 }{2 \epsilon H^2}  \gg  1 .  \label{ad-cond-metric-turn}
\ee
We may compute  the radius of curvature $\kappa$ of the turn by noticing that the unit vectors associated to a curve following the minimum of the potential $\psi = 0$ are given by
\be
T^{a} = (1,0), \qquad N^{a} = ( \Gamma , -1  ).
\ee
Plugging these expressions  back into  eq.~(\ref{time-deriv-T}) we find that the characteristic radius of curvature while the turn is at its pick, is given by:
\be
\kappa = \frac{1}{|\partial_{\chi} \Gamma|}  \simeq \frac{\Delta \chi}{|\Gamma_0|}.
\ee
This implies that $\eta_{\perp} = \dot \phi / H \kappa$ characterizing the turn is given by
\be
\eta_{\perp} \sim  \frac{ \sqrt{2 \epsilon} | \Gamma_0 |}{  \Delta \chi },  \label{eta-perp-in-terms-of-T-example-2}
\ee
whereas the speed of sound is found to be
\be
c_s^{-2}  \simeq 1 +  \frac{8 \epsilon H^2 \Gamma_0^2}{ \Delta \chi^2 M^2   }.
\ee
Notice that the only difference between this model and the previous one (where $a$ plays the role of $\Delta \chi$) is the appearance of $\Gamma_0$. By defining the following dimensionless parameter,
\be
\beta   \equiv   \frac{8 \epsilon H^2 \Gamma_0^2}{ \Delta \chi^2 M_{\psi}^2   },
\ee
we see that, in order to have large effects on the power spectrum due to the turn, we are required to have $\beta \sim 1$. Notice that one may satisfy this combination of parameters and still stay within the region of validity of the effective field theory, given by condition~(\ref{ad-cond-metric-turn}).

\subsubsection{Numerics}

We now  present our numerical analysis of this model. As in the example of Section \ref{sec:canonical-model}, we choose the following values for the potential parameters associated to the flat inflationary direction: $V_0 = 6.5 \times 10^{-9}$  and $V_{\phi} = - 5.4 \times 10^{-11} $ which in the absence of turns, imply $H \sim 1.4 \times 10^{-5}$,  $\epsilon \simeq - \eta \simeq 0.0033 $, and  $n_s \simeq 0.98$.  Notice that the rest of the parameters in charge of characterizing the multi-field turn are $\Delta \chi$, $\Gamma_0$ and $M_{\psi}$. These may be expressed in terms of $\alpha$, $\beta$ and $\Delta N_{\bot}$ introduced earlier as
\bea
M_\psi^2  & = & \alpha^2 \frac{2 \epsilon H^2}{\Delta \chi^2 }, \\
\Gamma_0^2 & = & \beta \frac{ \Delta \chi^2 M^2   } {8 \epsilon H^2 }, \\
\Delta \chi & = & \Delta N_{\perp} \sqrt{2 \epsilon}.
\eea 
Recall that the adiabaticity condition~(\ref{ad-cond-metric-turn}) is equivalent to $\alpha \gg 1$, and therefore we expect a poor matching between the full two-field theory and the EFT for small values of $\alpha$.

\begin{figure}[h!]
\begin{center}
\includegraphics[scale=0.75]{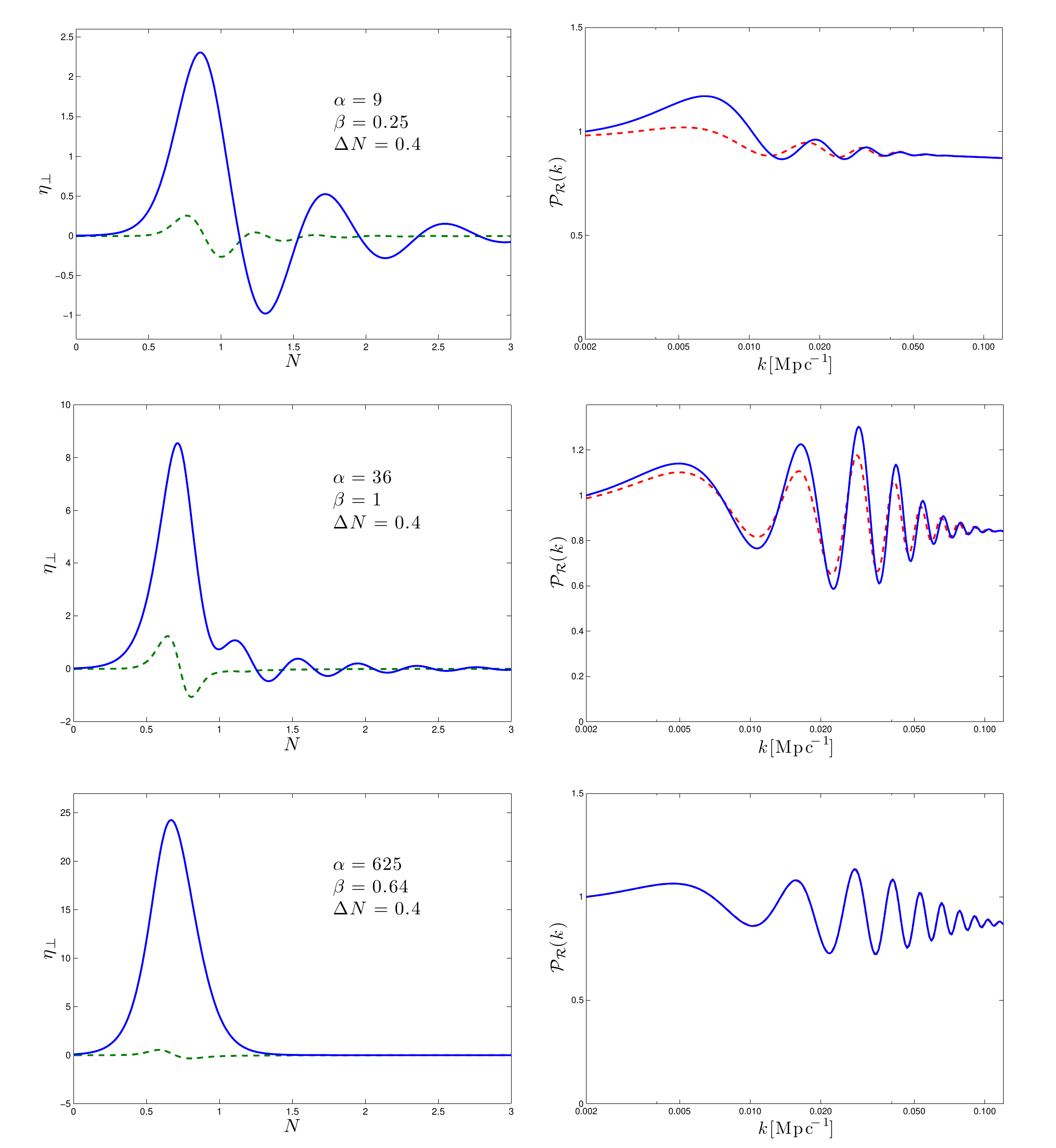}
\caption{\footnotesize The figure shows $\eta_{\perp}$ and $\eta_{||}$ (left panels) and the resulting power spectra (right panels) for three choices of parameters for the potential of our model 2. From top to bottom: $(\alpha , \beta) = (9, 0.25) $, $( 36, 1) $ and $(625 , 0.64) $. In the case of the left panels, the blue solid line corresponds to $\eta_{\perp}$ whereas the green dashed line corresponds to $\eta_{||}$. In the case of right panels, the blue solid line corresponds to the power spectrum for the full two-field model, whereas the red dashed line corresponds to the power spectrum deduced using the EFT. }
\label{examples-metric}
\end{center}
\end{figure}

Figure~\ref{examples-metric} shows three examples of turns for different values of the parameters $\alpha$ and $\beta$, but for a fixed value $\Delta N_{\perp} = 0.4$. 
The top panels correspond to the choice $\alpha = 9$ and $\beta = 0.25$. It may be seen that given that the value of $\alpha$ is relatively low, the adiabaticity condition is not satisfied. Consistent with the discussions of the previous sections, the dependence on time of $\eta_\perp$ is dominated by fluctuations with a period of oscillation determined by $T_M \sim 1/ M_{\psi}$, and the power spectrum obtained from the effective field theory (dashed red line) does not coincide with the one obtained from the complete two-field model (solid blue line). The middle panels correspond to the choice $\alpha = 36$ and $\beta = 1$. Here the adiabaticity condition is slightly improved while the speed of sound suffers a sizable change. Finally, the lower panels show the situation $\alpha = 625$ and $\beta = 0.64$. Here the adiabaticity condition is fully satisfied and the speed of sound becomes suppressed during a brief period of time. This is reflected by the excellent agreement between both power spectra, and their large features. 

It may be noticed that in the first two cases $(\alpha , \beta) = (9, 0.25) $ and $( 36, 1) $, there is a sizable variation of $\eta_{||}$. Just like in the case of our Model 1, this variation happens only during a brief period of time, and it is not enough to break the overall slow-roll behavior of the system. In addition, the value of $\epsilon$ is not affected considerably by the turn. In the last case $(\alpha , \beta) = (625, 0.64)$ the variation of $\eta_{||}$ is attenuated and the only background quantity varying considerably turns out to be $\eta_{\perp}$. This in turns makes $c_s$ to have large variations, therefore producing the large features observed in the resulting power spectrum.

\begin{figure}[h!]
\begin{center}
\includegraphics[scale=0.62]{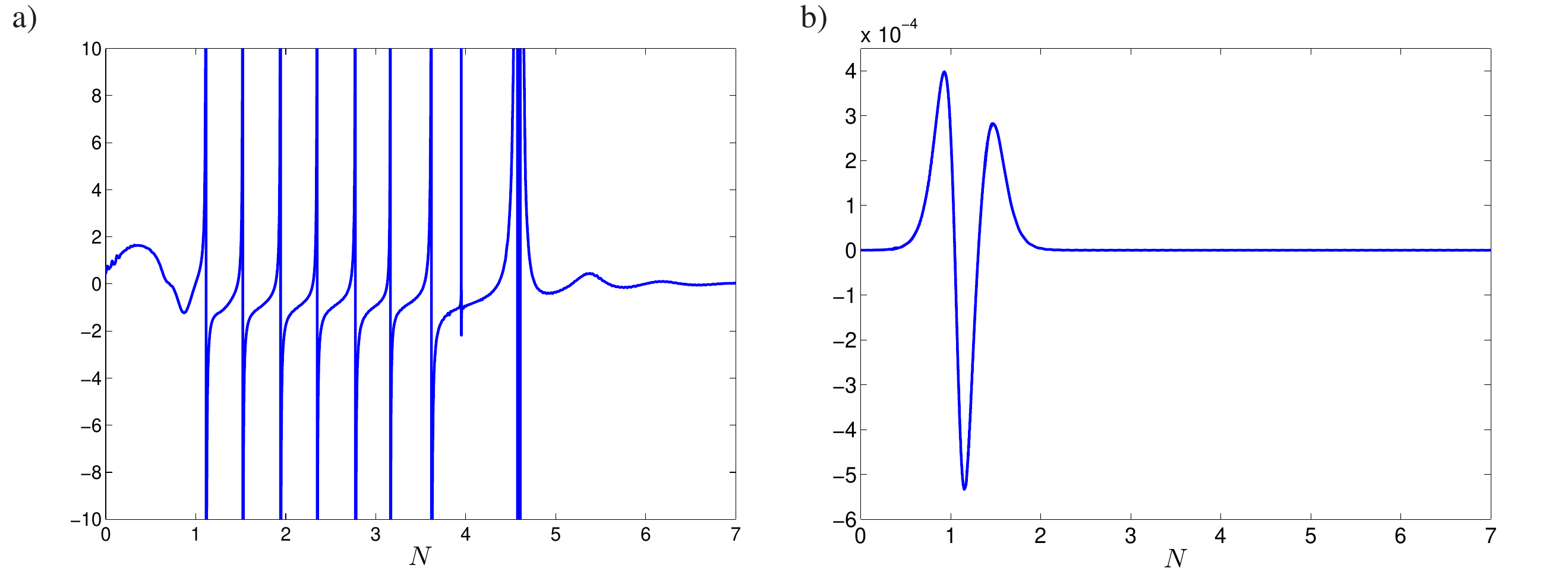}
\caption{\footnotesize The figure shows the the quantity $f(t)$ defined in eq.~(\ref{def-f}) for the cases $(\alpha, \beta) = (9, 0.25)$ and $ (625 , 0.64) $ respectively. This function assesses whether the adiabatic condition is being satisfied during a turn. (We have chosen to plot this function in terms of $e$-folds $N$ to facilitate its comparison with other quantities).}
\label{fig:adiabatic-cond}
\end{center}
\end{figure}

To finish, it is instructive to verify how does the rate of change of $\dot \theta$ evolves while the turn takes place. For this, we define
\be
f(t) = \frac{1}{M_{\rm eff}^2}  \frac{1}{\dot \theta} \frac{d^2}{dt^2} \dot \theta. \label{def-f}
\ee
According to our discussion in Section~\ref{sec:effective-theory}, the adiabaticity condition will be satisfied if $|f(t)| \ll 1$. Figure~\ref{fig:adiabatic-cond} shows $f$ as a function of $e$-fold $N$ for the cases $(\alpha, \beta) = (9, 0.25)$ and $ (625 , 0.64) $ respectively. It may be appreciated that indeed case $(\alpha, \beta) = (9, 0.25)$ is far from satisfying the adiabaticity condition, whereas the case $(\alpha, \beta) = (625, 0.64)$ satisfies it. 


\section{Conclusions} \label{sec:conclusions}
\setcounter{equation}{0}

In this work we have analyzed the dynamics of two-field models of inflation with large mass hierarchies. We have focussed our attention on the role that turning trajectories have on the evolution of perturbations. If the mass $M$ of the heavy field is much larger than the rate of expansion $H$, then the heavy field may be integrated out giving rise to a low energy effective field theory valid for curvature perturbations $\mathcal R$,  with a quadratic order action given by eq.~(\ref{total-action-R-effective}). At this order, all of the effects inherited from the heavy sector are reduced in the speed of sound $c_s$, which is given by eq.~(\ref{eq:speed-of-sound}). We found  that a good characterization of the validity of  the low energy effective theory is given by the following adiabaticity condition:
\be
\left| \frac{d}{d t} \ln \dot \theta  \right| \ll M_{\rm eff} , \label{conclusions:adiabatic-cond}
\ee
where $M_{\rm eff}$ is the effective mass of heavy modes. Our numerical analysis is consistent with this condition, and we find that several non-trivial effects are still significant within this allowed region of parameters. For instance, in Section~\ref{sec:sudden-turns}  we were able to provide two simple toy models for which the effective field theory remains fully trustable. In these examples large features are generated and appear superimposed on the primordial power spectrum of curvature perturbations (recall the examples of Figures~\ref{examples-potential} and \ref{examples-metric}). In addition, we verified that indeed as soon as the adiabaticity condition starts to fail, this is reflected in noticeable discrepancies in the power spectra predicted by both the complete two-dimensional model and the effective field theory (recall Figure~\ref{fig:adiabatic-cond}).

Our results contradict those of recent works regarding the validity of effective field theories obtained from multi-field inflation in various respects. For instance, in ref.~\cite{Shiu:2011qw} it is claimed that the effective field theory~(\ref{total-action-R-effective}) is only valid in the regime where turns are such that $| \dot \theta | \ll H$.\footnote{A similar claim is made in ref.~\cite{Peterson:2011yt}, where it is argued that heavy fields can only be integrated out consistently if the rate of turn satisfies $| \dot \theta | \ll H$.} The main argument made there is that the ratio $\eta_{\perp} =  \dot \theta / H$ corresponds to the coupling determining the kinetic energy transfer between the light curvature mode with the heavy fields. A large value of $\eta_{\perp}$ would therefore imply large transfer of energy from curvature perturbations to the heavy mode, exciting the heavy modes and rendering the effective field theory invalid.
However, as we have seen, this energy interchange between both modes may happen adiabatically without implying a breakdown of the effective field theory. Indeed, the heavy mode is receiving energy from the light degree of freedom at the same rate than it is giving it back, and therefore it is possible to have turns whereby the heavy-mode's high-frequency fluctuations stay suppressed (recall our discussion of Section~\ref{sec:effective-theory}).

One key point here is that even for large values of $\eta_{\perp} =  \dot \theta / H$ the heavy modes will not become easily excited unless they receive a sufficiently strong kick. For example, even if the turn is such that 
\be
\left| \frac{d}{d t} \ln \dot \theta  \right| \gtrsim H ,
\ee
the trajectory is necessarily subject to a large angular acceleration $| \ddot \theta | \gtrsim H | \dot \theta| $, momentarily violating slow-roll~\cite{Avgoustidis:2011em} but without exciting heavy modes. Moreover, as we have seen, if these accelerations are brief (as in our examples) slow-roll is only interrupted for a short period of time, and the system quickly goes back to the slow-roll attractor state (within an $e$-fold). During these transients, $\eta_{||}$ were typically found to have sizable variations in response to the turns, whereas $\epsilon$ was found to stay close to its suppressed value.\footnote{This was also found numerically in ref.~\cite{Achucarro:2010da}, and in ref.~\cite{Shiu:2011qw} an analytic argument was given to explain this effect.} The net effect of this process are oscillatory features in the power spectrum, with the frequency of the oscillation depending on how brief was the overall turn. While current observational constraints on features are still poor~\cite{Bridle:2003sa, TocchiniValentini:2004ht, Mukherjee:2005dc, Hunt:2007dn, Ichiki:2009xs, Peiris:2009wp, Hamann:2009bz, Hlozek:2011pc, Kumazaki:2011eb, Aich:2011qv, Meerburg:2011gd}, future data will certainly put strong constraints on primordial features, therefore improving our understanding of the role of UV-physics on the very early universe. 

Another important point of departure from previous works regards the procedure employed to integrate heavy fields. In the present work we have integrated out high energy fluctuations (heavy degrees of freedom) about the exact time-dependent background trajectory, offered by the homogeneous equations of motion of the system. This contrasts with other schemes~\cite{Yamaguchi:2005qm, BenDayan:2008dv, Gallego:2008qi} where heavy fields are taken care of at the action level, regardless of the background dynamics, by imposing that they locally minimize the inflationary potential (with the minima depending on the inflaton v.e.v.). In our present language this is equivalent to $V_N = 0$, implying that there are no turns at all (and therefore missing all of the interesting features we have studied so far). Although such a case corresponds to a genuine limit shared by a large family of multi-field potentials with mass hierarchies~\cite{Gallego:2011jm, Brizi:2009nn}, it misses the more general situation in which the inflationary trajectory meanders away from the locus of minima offered by the potential.

We should emphasize that our results are strictly valid only at linear order in the fluctuations. For a complete analysis one should examine the relevance of higher order interaction terms which could introduce important corrections when the speed of sound is suppressed ($c_s \ll 1$). This is similar to the case of DBI inflation~\cite{Silverstein:2003hf, Alishahiha:2004eh} where a suppressed speed of sound makes the perturbation theory to enter the strong-coupling regime~\cite{Cheung:2007st}.  In ref.~\cite{Achucarro:2012sm}  the full structure of effective field theories arising from two-field models with large mass hierarchies is studied, and other relevant constraints involving higher order terms are analyzed.

\section*{Acknowledgements}

We would like to thank Ana Ach\'ucarro, Cristiano Germani, Jinn-Ouk Gong, Sjoerd Hardeman and Subodh Patil for useful comments and discussions on the content of this work. 
This work was partially funded by Conicyt under the Fondecyt ``Initiation in Research" project 11090279 (GAP \& SC) and  by a  Leiden Huygens Fellowship (VA).
GAP wishes to thank King's College London, University of Cambridge (DAMTP), CPHT at the Ecole Polytechnique and the Lorentz Institute (Leiden) for their hospitality during the preparation of the manuscript.

\end{document}